\documentclass[sigconf]{acmart}

\AtBeginDocument{%
  \providecommand\BibTeX{{%
    \normalfont B\kern-0.5em{\scshape i\kern-0.25em b}\kern-0.8em\TeX}}}

\setcopyright{acmcopyright}
\copyrightyear{2024}
\acmYear{2024}
\acmDOI{10.1145/3613904.3642689}

\acmConference[CHI '24]{the CHI Conference on Human Factors in Computing Systems}{May 11--16,
  2024}{Honolulu, HI, USA}
\acmISBN{ 979-8-4007-0330-0/24/05}

\acmSubmissionID{4137}

\usepackage{subcaption} 
\usepackage{multirow}
\usepackage{soul}

\usepackage{colortbl}       
\usepackage{amsmath,bm}
\usepackage{enumitem}
\usepackage{xfrac}
\usepackage{float}
\usepackage{graphicx}
\usepackage[T1]{fontenc}

\setlist[itemize]{align=parleft,left=0pt..1.4em}
\setlist[enumerate]{align=parleft,left=0pt..1.4em}

\begin{document}

\title[Incremental XAI]{Incremental XAI: Memorable Understanding of AI with Incremental Explanations}

\author{Jessica Y. Bo}
\email{jbo@cs.toronto.edu}
\affiliation{%
  \institution{University of Toronto*}
  \authornote{Work was performed while at the National University of Singapore}
  \city{Toronto}
  \country{Canada}
}

\author{Pan Hao}
\email{panhao@u.nus.edu}
\affiliation{%
  \institution{National University of Singapore}
  \city{Singapore}
  \country{Singapore}
}

\author{Brian Y. Lim}
\email{brianlim@comp.nus.edu.sg}
\affiliation{%
  \institution{National University of Singapore}
  \city{Singapore}
  \country{Singapore}
}

\renewcommand{\shortauthors}{Bo, Pan, Lim}

\begin{abstract} 
  Many explainable AI (XAI) techniques strive for interpretability by providing concise salient information, such as sparse linear factors. However, users either only see inaccurate global explanations, or highly-varying local explanations. We propose to provide more detailed explanations by leveraging the human cognitive capacity to accumulate knowledge by incrementally receiving more details. Focusing on linear factor explanations (factors $\times$ values = outcome), we introduce Incremental XAI to automatically partition explanations for general and atypical instances by providing Base + Incremental factors to help users read and remember more faithful explanations. Memorability is improved by reusing base factors and reducing the number of factors shown in atypical cases. In modeling, formative, and summative user studies, we evaluated the faithfulness, memorability and understandability of Incremental XAI against baseline explanation methods. This work contributes towards more usable explanation that users can better ingrain to facilitate intuitive engagement with AI.
\end{abstract}



\maketitle

\section{Introduction}
As Artificial Intelligence (AI) systems become prevalent, it is paramount for explainable AI (XAI) to be developed to support their proper use and understanding~\cite{liao2021human, adadi2018peeking, rudin2022interpretable, abdul2018trends, lim2009and, arrieta2020explainable}.
Although much work has shown that XAI can improve satisfaction and trust \cite{liao2021human, narayanan2018humans, kocielnik2019will, lucic2020does, tsai2021exploring}, many studies have failed to demonstrate measurably improved understanding~\cite{poursabzi2021manipulating, kaur2022sensible}.
This requires users to ingrain AI explanations to quickly recall and apply the knowledge for decision making.
Providing short explanations like sparse linear models could help, but these would be too simplified to be faithful to the complex underlying AI decision, and mislead users~\cite{narayanan2018humans}.
In contrast, more expressive explanations may be more faithful, but can be challenging to read or recall~\cite{abdul2020cogam}, hindering their accessibility.
This is especially important for users to generalize their understanding of AI behavior for future scenarios~\cite{miller2019explanation}.
Hence, XAI needs to be sufficiently detailed, yet memorable to support effective understanding.

To help users develop a richer understanding of AI models,
instead of inundating users with complex explanations, we propose to explain \textit{incrementally}.
This is inspired from pedagogy, where students learn a concept gradually rather than all-at-once. For example, physics students learn about classical Newtonian mechanics for objects moving at common speeds, but later learn the theory of Special Relativity that describes objects at very high speeds with the Lorentz transformation. Understanding relativistic mechanics is very complicated and requires the foundational understanding of classical mechanics first.
Thus, we argue that users can eventually understand complex explanations and models, but they should be grounded on simpler explanations, and incrementally informed.

We propose a step towards elevating user understanding towards complex AI explanations with \textit{Incremental XAI}. This framework defines how to explain AI predictions from typical cases to outlier cases.
We investigate this for simple surrogate explanation models, specifically, sparse linear explanations that describe linear factors that multiply against feature values.
For example, a factor of $w^{\text{(Bathrooms)}} = \$17\text{k}$ explains the predicted price of a house based on \# Bathrooms by indicating that each bathroom adds \$17k, and that two bathrooms would contribute \$34k together.
We begin with 
 partitioning the dataset into subspaces, 
training a linear model in the majority (typical) subspace with Base factors, and
training a linear model in the minority (outlier) subspace with (Base + Incremental) factors.
To minimize new information to learn, we regularize the Incremental factors to be 0 when possible.
In our bathrooms example, the majority smaller houses could have a rate of \$17k/bathroom, while minority larger houses can have costlier bathrooms at \$17k + \$51k, perhaps due to luxury fittings.
We contribute:
\vspace{-0.5em}
\begin{itemize}
    \item The \textit{Incremental XAI} paradigm which enables gradual delivery of complex explanations, gaining the benefit of multiple lightweight explanations that achieves higher faithfulness.
    \item A tree-based incremental explanation using linear model trees, additive factors, and factor sparsity regularization. We also developed a tabular user interface to convey explanations incrementally, and contrasted this with baseline variants. 
    \item An evaluation of the faithfulness, usage, understanding, and memorability of Incremental explanations against Global, Subglobal, and Local baseline explanations in modeling, formative, and summative user studies.
    We compared Incremental explanations
    with Global explanations to evaluate if providing more detailed explanations based on category of cases (subspaces) helps understanding,
    with Subglobal explanations that are a baseline subspace model that explains each subspace independently, and
    with Local explanations since they are often singularly deployed primarily for instance-based explanations, but may be misused for general understanding.
    \item A discussion of how to generalize the Incremental XAI paradigm to other applications and AI explanations.
\end{itemize}

\section{Background and Related Work}
Explainable AI (XAI) remains problematic for human interpretation due to the inaccuracy of overly simplified methods.
Here, we give a primer on XAI and and their cognitive demands, techniques to mitigate cognitive load, the need to provide multiple explanations, and XAI techniques partitioned into subspaces to improve accuracy.

\subsection{Surrogate explanations of AI}
Explanations of AI can improve user understanding by providing surrogate explanations of accurate AI models, or making "glassbox" models that are intrinsically interpretable.
However, the latter approach may have limited accuracy since these models tend to be overly simple. 
Instead, we focus on providing surrogate explanation models that approximate complex AI models that retains the use of accurate AI models while explaining with some unfaithfulness.

Miller~\cite{miller2019explanation} identified two goals for explanations in AI:
i) to select a small set of causes for an observation~\cite{lombrozo2006structure}, and
ii) generalize observations into a conceptual model to predict and control future cases~\cite{heider2013psychology}.
Wang et al. identified other reasoning processes that XAI should support~\cite{wang2019designing}.
Our research objective is to improve XAI techniques to better support the second goal of a generalized understanding.
This requires explanations to be intuitive and memorable so that users can rapidly apply their knowledge to anticipate the AI model's behavior in future settings.
Global explanations provide a suitable basis to support this goal.
They answer the question ``\textit{How} does the AI model make predictions?''
Techniques include explaining the AI decision in terms of linear factors~\cite{poursabzi2021manipulating}, nonlinear partial dependence plots~\cite{krause2016interacting} and generalized additive models~\cite{abdul2020cogam, caruana2015intelligible}, and decision trees~\cite{quinlan1986induction} and rules~\cite{lakkaraju2016interpretable,letham2015interpretable, ribeiro2018anchors}.

To support the former goal of explaining causes for an individual case, instance explanations are provided instead.
These answer the question: ``\textit{Why} did the AI model make this prediction?''
Techniques include feature attributions~\cite{lundberg2017unified,bach2015pixel,sundararajan2017axiomatic,datta2016algorithmic}, and counterfactual explanations~\cite{byrne2019counterfactuals,wachter2017counterfactual}.
An instance explanation only explains the decision for a target instance and may provide a different explanation for another instance; thus it may not generalize to multiple instances.
To overcome this, Ribeiro et al. proposed local explanations to train explainer models on instances similar to the target instance of interest~\cite{ribeiro2016should}.
Since these explanations focus on narrower sets of instances, they are more faithful to the underlying AI being explained, but require users to remember many models for dissimilar instances. 
Given their ubiquity in XAI practice, we include them in our investigations.
In this work, we aim to provide explanations that are memorable like Global explanations and faithful like Local explanations, investigated Subglobal explanations that balance between the two, and proposed Incremental explanations that improve the memorability of Subglobal explanations.

\subsection{Cognitive demands of AI Explanations}
Although XAI aims to improve user understanding, they are not necessarily easy for users to interpret~\cite{lim2023diagrammatization}.
High cognitive load harms user experience and the effectiveness of AI explanations \cite{poursabzi2021manipulating, lage2019human}. 
This is often measured by the number of attributes used in explanations~\cite{doshi2017towards} or the nonlinearity of visualizations~\cite{abdul2020cogam}.
Indeed, people consider simpler explanations as more probable than those with more clauses~\cite{lombrozo2007simplicity}, but oversimplifying explanations will erode trust in XAI~\cite{nourani2022importance, yin2019understanding}.
Explanations need to be delivered at the right level of cognitive effort to be effective \cite{gregor1999explanations, springer2020progressive, langer2021we, kim2023help}. 

A simple method to get users to understand explanations is to prompt users to think when reading explanations~\cite{buccinca2021trust}, but this does not ensure deep learning and understanding or make explanations less cognitively demanding.
Several techniques have been proposed to reduce cognitive load.
The most common is to do feature reduction to limit the number of attributes shown to users. This can be accomplished with sparsity regularization~\cite{narayanan2018humans} and constraining explanations to use integer coefficients instead of real numbers~\cite{ustun2016supersparse}.
However, this limits the expressiveness of explanations that users could consume.
Another approach is to simplify more sophisticated visual explanations, such as nonlinear line graphs. Cognitive-optimized GAM (COGAM) balances cognitive load and accuracy by quantifying the visual cognitive chunks in line chart explanations, and providing a hybrid explanation with sparse linear factors and less curvy line charts~\cite{abdul2020cogam}. 
However, these approaches only optimize one explanation at a time, but neglects the human cognitive capacity to accumulate knowledge.

\subsection{Providing multiple AI explanations}
Accurate understanding of an AI system requires detailed knowledge of its parameters and non-linear decisions, yet explanations need to be simple for easy comprehended. To avoid information overload, detailed explanations can be provided on demand~\cite{lim2011design} or with progressive disclosure~\cite{springer2020progressive}.
Users have various demands for explanations~\cite{lim2009assessing}, diverse usage strategies of explanations~\cite{lim2011design}, and use multifaceted explanations to understand AI decisions~\cite{lim2013evaluating, wang2022interpretable}. 

Hence, instead of considering XAI interpretation as independent interactions, it should be considered as sequentially dependent accumulation of knowledge (e.g., dialogic~\cite{miller2019explanation}).
People use explanations to build a mental model of the AI, so successful explanations can be measured through the goodness of the learned mental model \cite{liao2021human, wang2023watch}. Mental models play a key role in human-AI interactions \cite{bansal2019beyond}, but can be formed poorly without intervention \cite{druce2021brittle}.
In this work, we propose a new paradigm of providing explanations \textit{incrementally} by ensuring that the shallower, simpler and explanations can smoothly transition into deeper, more detailed ones. This leverages the human ability for cumulative learning~\cite{shuell1986cognitive}, and allows users to understand how the explanations relate~\cite{zhang2022towards} at different levels.

\begin{figure*}[h!]
    \centering
    \includegraphics[width=0.7\textwidth]{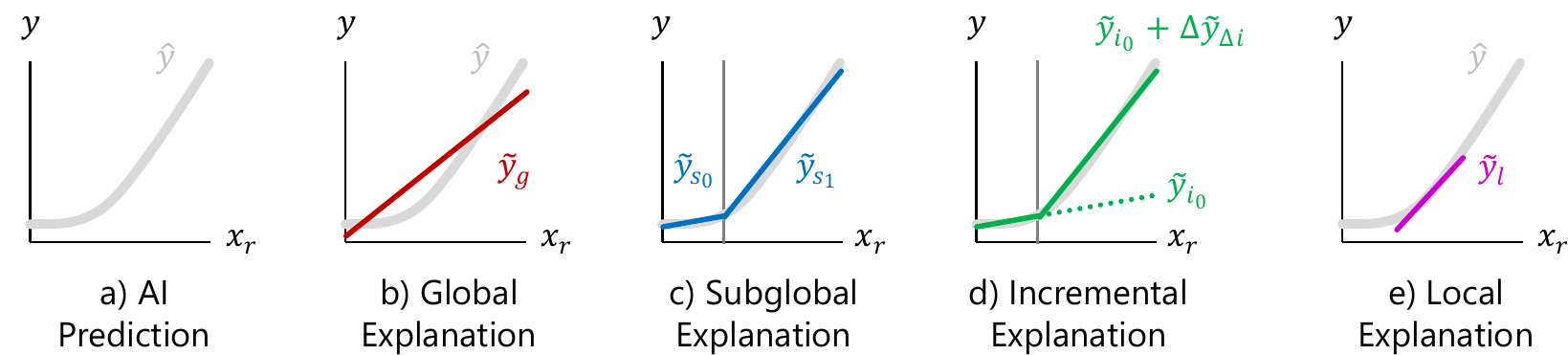}
    \vspace{-0.1cm}
    \caption{
    Conceptual examples of XAI types with univariate (1D) data shown for simplicity; see Fig. \ref{fig:xai_types_2d} for 2D multivariate examples with real data.
    a) Original AI System predicts output $\hat{y}$ non-linearly with respect to attribute $x_r$.
    b) Global explainer that approximates $\hat{y}$ with a linear equation $\tilde{y}_g \propto x_r$.
    c) Subglobal explainer increases faithfulness by segmenting along $x_r$ to provide multiple linear explanations $\tilde{y}_{s_1} \propto x_r, x_r < \chi_r$ and $\tilde{y}_{s_2} \propto x_r, x_r \geq \chi_r$.
    d) Incremental explainer that is similar to Subglobal, but first explains with a linear model $\tilde{y}_{i_0} \propto x_r$ the contiguous majority of instances (in this case, $x_r < \chi_r$), then explains outlier instances ($x_r \geq \chi_r$) with an additive linear model $\tilde{y}_{i_0} + \Delta\tilde{y}_{\Delta i}$.
    e) Local explanation explains each instance with a linear equation $\tilde{y}_l \propto x_r$ based on neighboring instances.
    Multiple local explanations are needed to represent the full input space.
    }
    \label{fig:xai_types_1d}
\end{figure*}

\bgroup
\def\arraystretch{1.1}
\begin{table*}[t]
    \caption{
    Comparison of linear explanation models with varying faithfulness and memorability due to the \# factors used, which affects their expressiveness and the \# terms for the user to remember.
    The AI System typically has too many factors to be interpretable, while sparse linear explanations consider a sparse set of $n$ factors.
    A Global explanation and Local explanation have 1 set of $n$ factors, but the latter requires many $N$ explanations to understand all use cases cumulatively.
    Subglobal explanations split the instances into $k$ subspaces, needing $kn$ factors to explain fully.
    Incremental explanations similarly can explain the same $k$ subspaces, but reuses some factors, and can omit negligible changes, thus it has $\leq kn$ factors.
    }
    \label{tab:xai_comparisons}
    \vspace{-0.1cm}
    \begin{tabular}{rccccc}
    \toprule
    \textbf{}                        & \multirow{2}{*}{\begin{tabular}[c]{@{}c@{}}Prediction  Model\end{tabular}} & \multicolumn{4}{c}{Explainer Models}                                                                     \\ \cline{3-6} 
    \textbf{}                        &                                                                              & Global     & Subglobal                     & \textbf{Incremental}   & Local \\ \hline
    \# Factors & $\gg n$ & $n$ & $kn, k\geq2$ & $\le kn$ & $Nn, N \gg n$ \\
    Faithfulness                     & Self                                                                         & Low        & Med                           & Med           & High \\
    \multicolumn{1}{l}{Memorability} & Low                                                                          & High       & Med-Low                       & Med-High      & Low                                         \\ \bottomrule
    \end{tabular}
\end{table*}
\egroup

\subsection{Subspace-based XAI techniques}
Several XAI techniques have been developed to address the shortcomings of global explanations beingf too coarse and local explanations being too narrow. We discuss methods that divide instances into subspaces and explain each subspace separately. Methods are based on trees, rules, or aggregation.

Model agnostic multilevel explanations (MAME)~\cite{natesan2020model} provides an explanation tree with weights at each node, representing a progression from a global explanation at the root to local explanations at the leaves. However, their method does not enforce stability between each linear model, so it would be difficult for users to learn each sub-explanation incrementally.
Model Understanding through Subspace Explanations (MUSE)~\cite{lakkaraju2019faithful} provide decision sets for different subspaces by simultaneously optimizing for faithfulness and rule compactness, but the explanations are in terms of rules unlike our factors-based format, and the attributes are not necessarily the same for each subspace, thus not consistent.
Equi-explanation Maps~\cite{chowdhury2022equi} divide a feature space into hyper-cuboid subspaces (i.e., defined within min/max ranges for specific attributes) that are consistent, and explain each subspace with linear classifiers, but its boundary definitions are much more complex.
Submodular Pick LIME~\cite{ribeiro2016should} leverages instance-based local LIME explanations to provide a global explanation by picking diverse LIME explanations that have high non-redundant coverage. This aims to limit the total number of Local explanations needed to achieve global understanding.
GLObal to loCAL eXplainer (GlocalX)~\cite{setzu2021glocalx} iteratively merges local decision rules into global explanations to provide a smooth pathway from detailed local explanations to more general global explanations and vice versa. This was demonstrated for rule-based explanations of classification, unlike our prediction regression task with linear factors. 
Sparse LInear Subset Explanations (SLISE)~\cite{bjorklund2019sparse} is a robust regression method that finds the largest subset in the data and trains a sparse linear model. Our method could use this to learn the base model of the Incremental explanation.
SLISEMAP~\cite{bjorklund2023slisemap} extends SLISE to group instances into clusters based on the similarity of their local explanations. This involves dimensionality reduction, so the resulting dimensions are not explicitly interpretable.

All these methods aim to explain each subspace faithfully, but neglect to account for users having to remember or relate across subspaces. Thus the inter-subspace consistency is low. In our work, we focus on first providing a base explanation for a majority subspace, and explain remaining smaller subspaces as incrementally different from the base. This new requirement stems from usability needs of XAI that the prior works neglect.

\section{Technical Approach}

\begin{figure*}[h]
    \centering
    \includegraphics[width=0.85\textwidth]{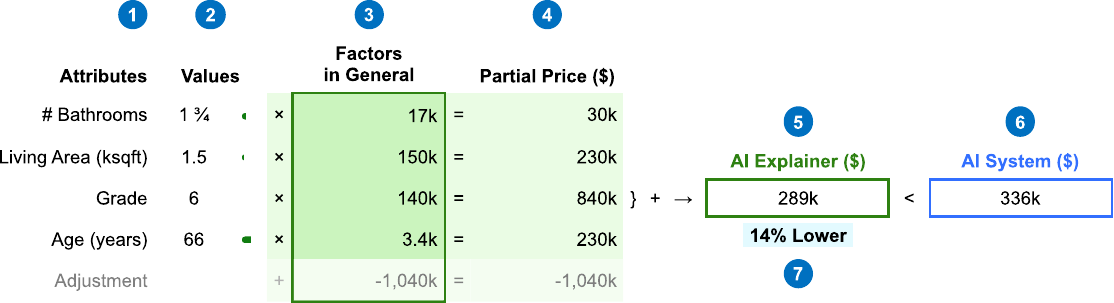}
    \vspace{-0.1cm}
    \caption{
    User interface (UI) of AI System with Global explanation showing:
    1) attributes used for prediction,
    2) their values $x^{(r)}$ for the given instance,
    3) factors $w_r$ that the explainer multiplies with values,
    4) partial contributions $\tilde{y}_i = w^{(r)} x^{(r)}$ of each attribute,
    5) output estimation $\tilde{y} = \sum_{r}{\tilde{y}^{(r)}}$ from the AI Explainer,
    6) prediction $\hat{y}$ from the AI System, with inequality indication ($<$ in this case), and
    7) indicator of how different the AI Explainer estimation is from the AI System prediction.
    Factors are the same for all instances and do no change.
    Different information may be hidden under various test conditions. 
    }
    \vspace{-0.1cm}
    \label{fig:ui_global}
\end{figure*}

We first describe baseline explanation approaches using sparse linear factors, then articulate our Incremental explanation approach. 
An AI System's prediction $\hat{y}$ is typically generated from many input attributes (features) $\bm{x}$, and the AI's decision may change non-linearly with each attribute (e.g., price can increase exponentially with living area of a house). 
Sparse linear models provide simple explanations by articulating only a few important attributes (hence sparse), and indicating how each attribute influences the prediction.
These are typically presented as \textit{feature attributions}, i.e., positive or negative numbers indicating the direction and magnitude of the influence.
However, attributions are not particularly easy to track or interpret, since they vary inconsistently for different instances.

Instead, like Poursabzi-Sangdeh et al.~\cite{poursabzi2021manipulating}, we focus on sparse linear factor explanations that compute the feature attribution of the $r$th attribute $y^{(r)}$ as a multiplication of a factor weight $w^{(r)}$ and the attribute value $x^{(r)}$, i.e., $\tilde{y}^{(r)} = w^{(r)} x^{(r)}$.
For example, consider a house with 1\sfrac{3}{4} bathrooms, i.e., $x^{(1)} = 1.75$. $w^{(1)} = 17$k means that each increase in one bathroom costs \$17k more, so the \# Bathroom contributes $\$17\text{k} \times 1.75 = \$30\text{k}$ to the total house price.
Users can apply these factors to another instances to calculate how the AI Explainer would estimate the AI prediction for that instance.
For example, a house with 3 bathrooms would have its \# Bathrooms contribute $\$17\text{k} \times 3 = \$51\text{k}$ to its price.
Sparse linear factors can be applied broadly to all instances (Global explanation), semi-broadly to groups of instances (Subglobal explanation), or to individual instances (Local explanation). We introduce each of these explanation methods and then describe our approach for the Incremental explanation that extends Subglobal. Each type has varying \textit{faithfulness} to the AI prediction and \textit{memorability} for users to recall the factors, which we summarize in Table \ref{tab:xai_comparisons} and illustrate conceptually in a 1-dimensional example in Fig. \ref{fig:xai_types_1d}. We refer to these explanation variants as XAI types.

\subsection{Global explanation}

The simplest explainer uses a single linear factor model with one set of factors to explain for all instances.

\begin{equation}
\label{eq:global}
\tilde{y}_g = \sum_{r}{w_g^{(r)} x^{(r)}}
\end{equation}

where $x^{(r)}$ is the $r$th feature value of the instance with $x^{(0)} = 1$, 
$w_g^{(r)}$ is the explanation factor for that feature with $w_g^{(0)}$ as the bias term,
and $\tilde{y}_g$ is the estimated AI prediction.
The Global explanation is trained by fitting a linear regression model on the whole training dataset with mean squared error (MSE) as the training loss against the AI model's prediction not the ground truth.
Fig. \ref{fig:ui_global} shows our user interface (UI) implementation of a Global explanation of how an AI System predicts the price of a house based on 4 attributes, and the bias term which we name as "adjustment".

\subsection{Subglobal explanation}

\begin{figure*}[h]
    \centering
    \begin{subfigure}{0.85\textwidth}
        \includegraphics[clip,width=\columnwidth]{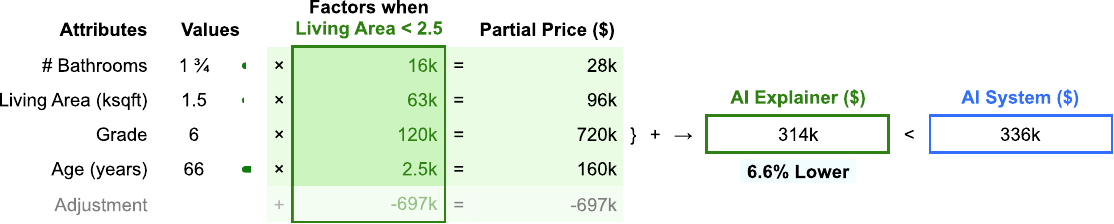}
        \vspace{.00cm}
    \end{subfigure}
    \begin{subfigure}{0.85\textwidth}
        \includegraphics[clip,width=\columnwidth]{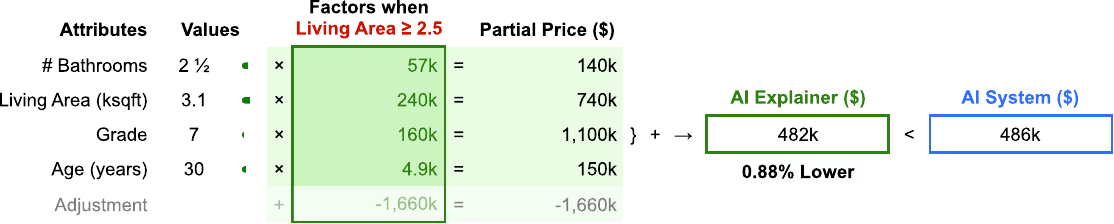}
    \end{subfigure}
    \caption{User interface (UI) of Subglobal explanations for a typical instance (top), and an outlier instance (bottom). 
    Factors are different for each subspace but apply in a fixed way to any instance in each subspace.
    For example, while small houses with Living Area < 2.5 ksqft have each bathroom being worth \$16k, larger houses have much costlier bathrooms at \$57k.
    }
    \vspace{-0.2cm}
    \label{fig:ui_subglobal}
\end{figure*}

\begin{figure*}[h]
    \centering
    \begin{subfigure}{0.95\textwidth}
        \includegraphics[clip,width=\columnwidth]{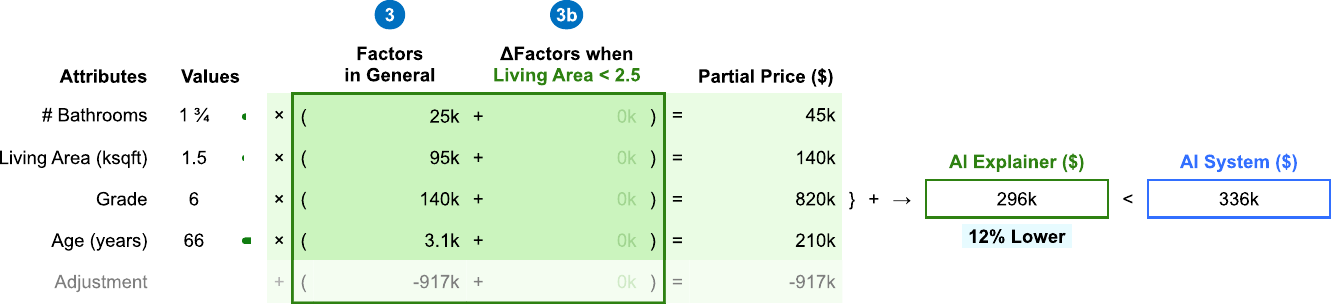}
        \vspace{.1cm}
    \end{subfigure}
    \begin{subfigure}{0.95\textwidth}
        \includegraphics[clip,width=\columnwidth]{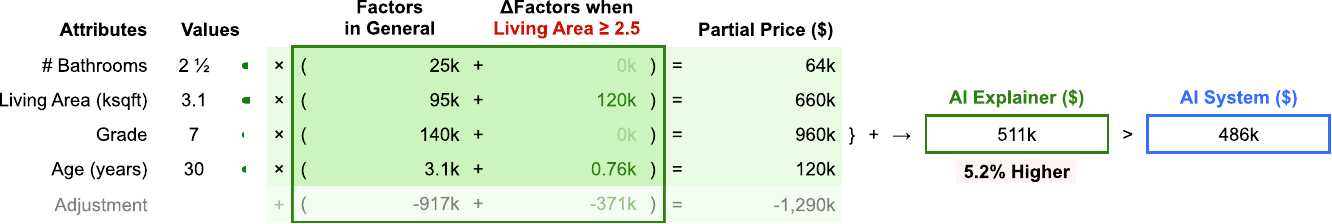}
    \end{subfigure}
    \caption{
    User interface (UI) of Incremental explanation for an instance in the typical subspace with Living Area < 2.5 (top), and an outlier instance in the minority subspace with Living Area $\geq$ 2.5 ksqft (bottom).
    Factors are different for each subspace to fit them accurately.
    Unlike Subglobal explanations, an additional column (3b) is used to show how factors are incrementally different for the outlier cases. The main factors (3) are the same for both subspaces.
    For example, while smaller houses have a modest rate of price increase per living area (\$95k/ksqft), larger houses have a rate that is \$120k/ksqft higher (\$215k/ksqft).
    }
    \vspace{-0.2cm}
    \label{fig:ui_incremental}
\end{figure*}

While the Global explanation is simple for users to understand, its small number of factors limits its expressiveness, so it may not be very faithful to the AI System predictions, i.e., $\tilde{y}_g$ is not close to $\hat{y}$. 
Instead of adding more complexities for users to interpret, the explanation faithfulness can be increased by partitioning instances into multiple subspaces. Each subspace is then modeled with a separate sparse linear factor explanation. 
We constrain explanations of each subspace to have the same attributes, and enforce the partition based on binary univariate rules, i.e., inequality on one attribute (e.g., $x_2 \geq 2.5$).
Thus, a Subglobal explanation has the form
\begin{equation}
    \label{eq:subglobal}
    \tilde{y}_s = \sum_{\varsigma}\sum_r{[x \in s_\varsigma] w_{s\varsigma}^{(r)} x^{(r)}} = 
    \begin{cases}
        \sum_r{w_{s0}^{(r)} x^{(r)}}, & \text{if } x \in s_0 \\
        \sum_r{w_{s1}^{(r)} x^{(r)}}, & \text{if } x \in s_1 \\
        ...
    \end{cases}
\end{equation}
where $\bm{w}_{s\varsigma}$ is the weights of the $\varsigma$th subspace explanation model,
$[\cdot]$ is the Iverson bracket that is 1 if its expression is true or 0 otherwise, 
and $S_\varsigma$ is the set of instances in subspace $\varsigma$.
Eq. \ref{eq:subglobal} shows that each subspace has different weights (factors) which it applies to instances within its boundaries.

Training the Subglobal explanation model requires learning the partition boundaries of the subspaces, and the weights of each subspace model. 
We achieve this by training a linear model tree~\cite{cerliani2022lineartree} on the whole training dataset with MSE for the training loss.
Such trees are different from common classification decision trees that predict a probability distribution of categorical labels $p(\hat{y})$ at leaves, or regression decision trees that predict a scalar number $\hat{y}$ at the leaves. 
Instead, linear model trees predict a linear regression model $\bm{w}_\varsigma$ at each leaf, where each leaf represents a subspace $S_\varsigma$. 
During training, for each branch in the decision tree, the training algorithm iterates through all features and possible splits, training a linear model for each subspace ($s_{\leq}$ and $s_{\geq}$), measuring the combined loss for both models, and choosing the split with the lowest combined loss.
We then assign the majority subspace with the larger dataset as ``typical'' and minority one as ``outliers'' (although this can be flexibly adapted to fix user preferences or standard conventions).
Though linear model trees are not a novel technique, they are seldom used in explainable AI, and we extend it for Incremental explanations for our technical contribution, described in the next subsection.

Fig. \ref{fig:ui_subglobal} shows our UI of Subglobal explanations with two subspaces: typical (Living Area < 2.5 ksqft) and minority outlier (otherwise).
For simplicity, we specifically train a decision stump (one branch).
Each subspace is defined with simple univariate decision boundaries that are easy to interpret.
Note that training a logistic regression or linear support vector machines (SVM) would lead to less interpretable decision boundaries, e.g., 
``$16 (\text{\# Bathrooms}) + 120 (\text{Grade}) < 697$'',
while training on a decision tree would have a more interpretable rule, e.g., ``\# Bathrooms $\geq$ 5 and Grade $<$ 5''.

\subsection{Incremental explanation}

\begin{figure*}[h]
    \centering
    \includegraphics[width=0.85\textwidth]{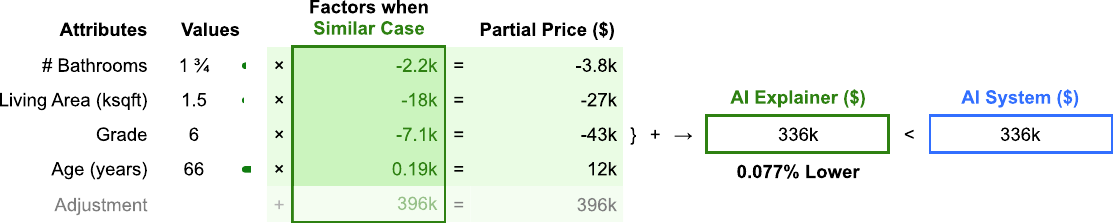}
    \caption{
    User interface (UI) of the Local explanation of an instance.
    Factors are specific to this and similar instances, and will be different for other instances.
    For example, for houses similar to the one shown, each increase in Grade decreases the house price by \$7.1k, but this may not be the case for other houses that have very different attributes.
    }
    \label{fig:ui_local}
\end{figure*}

While Subglobal is more faithful to the AI System than Global, this comes at a cost of the user having to read and remember more factors.
The factors are not necessarily consistent between subspaces too, so users would have to interpret them independently.
To improve memorability, we propose Incremental explanations that provide general factors for the majority, typical subspace and an incremented factors for special, outlier subspaces. We describe our approach for two subspaces, but it can be extended for multiple subspaces. 
We define an Incremental explanation as:
\begin{equation}
\begin{aligned}
\label{eq:incremental}
    \tilde{y}_i & = 
    \sum_{r}{\left(w_{i0}^{(r)} + \sum_{\varsigma > 0}{[x \in s_\varsigma] \Delta w_{\Delta i\varsigma}^{(r)}} \right)} x^{(r)} \\
    & = 
    \begin{cases}
        \sum_r{w_{i0}^{(r)} x^{(r)}}, & \text{if } \bm{x} \in s_0 \\
        \sum_r{\left(w_{i0}^{(r)} + \Delta w_{\Delta i1}^{(r)} \right) x^{(r)}}, & \text{if } \bm{x} \in s_1 \\
        ...
    \end{cases}
    \end{aligned}
\end{equation}
where $w_{i0}$ is the base factors of the general explanation model,
${\Delta w}_{\Delta i\varsigma}$ is the incremental factors of the $\varsigma$th special subspace explanation model,
$[\cdot]$ is the Iverson bracket that is 1 if its expression is true or 0 otherwise, 
and $S_\varsigma$ is the set of instances in subspace $\varsigma$.
Eq. \ref{eq:incremental} shows that while the typical subspace has Base factors, all other subspaces have factors defined as an additive Incremental adjustment on the Base factors.

We train the Incremental explanation model in a similar manner as for Subglobal explanations but constrain a dependency in factors across subspaces, i.e., $w_{i\varsigma}^{(r)} = w_{i0}^{(r)} + \Delta w_{\Delta i\varsigma}^{(r)}$.
Furthermore, to reduce the number of terms to remember, we aim to keep most incremental weights ${\Delta w}_{\Delta i\varsigma}$ to be zero. This is achieved by adding a sparsity L1 regularization to the original MSE training loss, i.e.,
\begin{equation}
    \label{eq:training}
    L(\tilde{y}_i, \hat{y}) = \sum_k{\left(\tilde{y}_i^{(k)} - \hat{y}^{(k)}\right)^2} + \lambda \sum_{\varsigma}\sum_r{\lvert{\Delta w}_{\Delta i\varsigma}\rvert}
\end{equation}
where $k$ indexes the training instances, and $\lambda$ is the regularization hyperparameter.
This sparsity regularization makes incremental factors easier to remember, but
this trades-off accuracy, so we hypothesize that the Incremental explanation $\tilde{y}_i$ is less faithful than the Subglobal explanation $\tilde{y}_s$.
The training algorithm is similar as in Subglobal explanations, but with non-independent parameters for the linear models, extended loss function, and unified optimization of both set of factors.
For each candidate split, we set the majority subspace as "typical" assigning the base factors\footnote{Note that the base factors only represent those of the "typical" instances (majority subspace), not of the Global explanation model; this is not the same due to the constraint to minimize incremental factors which will shift the base weights during training.} 
$w_{i0}$ to it, and specifying Incremental factors $\Delta w_{i\varsigma}$ in other minority "outlier" subspace.
Training is performed using gradient descent.

Since the partitioning of subspaces is similar in both Incremental and Subglobal explanations, we keep them the same in the modeling and user studies to avoid partitioning being a confounder.
Fig. \ref{fig:ui_incremental} shows our UI implementation of an Incremental explanation for cases in the typical and outlier subspaces.

\subsection{Local explanation}

We have described sparse linear factors to explain multiple instances, but they can also be used to explain individual instances. Local explanations, such as LIME~\cite{ribeiro2016should} and SHAP~\cite{lundberg2017unified}, are popular XAI techniques to explain AI decisions on a target instance by training a linear regression model from a dataset of instances that are local (similar) to the target instance. 
Explaining other instances require retraining other explanation models locally around those instances, and the explanations are not necessarily similar to one another. Hence, Local explanations are faithful to the AI prediction only to instances that are similar, and not globally or subglobally. Consequently, users would need to view many Local explanations to have an overview of the AI behavior across all instances.
Although local explanations are not designed for general understanding, their ubiquity encourages their misuse for this objective.
We hypothesize that this makes it very difficult for users to estimate how the AI System would predict for new instances
or estimate general factors from the inconsistent factors of each instance.
We define local linear factor explanations around a target instance $x_l$ as:
\begin{equation}
    \label{eq:local}
    \tilde{y}_l = \sum_{r}{w_l^{(r)} x^{(r)}}, \forall x \approx x_l
\end{equation}
where $x$ is the instance being explained that is similar to $x_l$, 
$\bm{w}_l$ is the weights of the model (factors) with $w_l^{(0)}$ as the bias term,
and $\tilde{y}_l$ is the estimation of the local model.
We implemented the Local explanation with LIME~\cite{ribeiro2016should}.
Fig. \ref{fig:ui_local} shows our UI implementation of a Local explanation around an instance.

\section{Evaluation}

We evaluated Incremental explanations against baseline explanations (Global, Subglobal, Local) across multiple studies to investigate:
i) faithfulness to estimate the AI prediction in a modeling study, 
ii) usage strategies and outcomes to interpret AI decisions in a qualitative formative user study, and 
iii) impact on decision duration, explanation recall, and AI decision understanding in a quantitative summative user study.

\subsection{Modeling Study}

We conducted a modeling study to evaluate how faithfully each explanation model estimates the AI. We evaluate on three datasets, and our approach can further generalize since we are using standard machine learning processes. We describe the dataset preparation, methods to train and test the models, and evaluation results.

\subsubsection{Applications and datasets}

We evaluated the sparse linear factor explanations on a regression prediction task, since the predictions remain linear, unlike classification that would have tapered effects at high or low probabilities (e.g., logistic regression).
Like Poursabzi-Sangdeh et al. ~\cite{poursabzi2021manipulating}, we evaluated on a housing price dataset due to the simplicity of the application scenario that most users can readily understand and appreciate. However, we chose not to reuse their NYC dataset since, surprisingly, a linear global model is sufficient to predict prices highly accurately. However, real-world datasets tend to be more complex, and require nonlinear models. Hence, we used the "House Sales in King County, USA" dataset~\cite{kingcounty} with 21,613 instances to predict the price of houses with 22 features. Prices ranged from \$72k to \$7.7M (Median = \$452k). 
We performed feature selection to obtain four features (\# Bathrooms, Living area, Grade, Age) to limit the cognitive load for users.

For generality, we further evaluate on two additional datasets: Heart Disease~\cite{misc_heart_disease_45} with 1025 instances to predict heart disease using 14 common features, and Auto MPG~\cite{misc_auto_mpg_9} with 398 instances to predict the miles per gallon fuel efficiency using 7 features.
Although the prediction task for heart disease is to classify whether a patient has heart disease, the predictor model produces a numeric confidence that can be interpreted as a continuous risk score. We train subsequent explainer models as a regression task to predict the risk score of the predictor model. 
For the heart disease dataset, to support human interpretability of the explainer models, we performed feature selection to obtain four features (Age, Resting blood pressure, Cholesterol, Max heart rate).
Similarly, for the Auto MPG dataset, we performed feature selection to obtain four features (Cylinders, Displacement, Horsepower, Weight).

For simplicity, we partitioned each dataset into two subspaces and set the same rule boundary for both Subglobal and Incremental explanations.
The optimal partitions were at Living Area $\geq$ 2.5 ksqft for House sales, Age $\geq$ 58 years for Heart disease; and Horsepower $\geq$ 92W for Auto MPG.

\subsubsection{Results on performance of AI prediction models}
For each dataset, we trained a random forest regressor (House Sales, Auto MPG) or classifier (Heart Disease) ~\cite{breiman2001random} as an AI prediction model on a training set of 80\% instances, and evaluated on a heldout test set of 20\% instances.
We then trained the four explanation (XAI) types to explain all instances in the test set. For Local explanation, we averaged the performance across all instances.
Using the training set, we performed 5-fold cross validation in all our analyses and report the mean and standard deviation of the validation performance averaged across folds; see Table \ref{tab:kfolds}.
We report the performance on the heldout test dataset --
House Sales: mean absolute error (MAE) = \$139k and $R^2$ = 0.67;
Heart Disease: accuracy = 86\% and test AUC = 0.86;
Auto MPG: MAE = 3.12 mpg and $R^2$ = 0.71.

\begin{table*}[t]
\small
\caption{Modeling results from 5-fold cross-validation of AI performance and XAI faithfulness across three datasets showing mean $\pm$ standard deviation.
AI performance indicates when an explainer is trained on the \textit{ground truth} dataset as a glassbox interpretable model.
XAI unfaithfulness evaluates each explainer as a surrogate explanation with respect to the AI Model.
Except for AI performance for Heart Disease that is measured as \% Accuracy, all other metrics are MAE, where smaller is better.
}

    \vspace{-0.25cm}
\begin{tabular}{llccccc}
\hline
\multicolumn{2}{l}{\textbf{House Sales}}   & \multirow{2}{*}{\begin{tabular}[c]{@{}c@{}}AI\\ Model\end{tabular}} & \multicolumn{4}{c}{XAI types}                                    \\ \cline{4-7} 
Subspace     & Metric (MAE \$k)              &                                                                     & Global         & Subglobal      & Incremental    & Local         \\ \hline
Combined     & AI Performance (inv)          & $132.5 \pm 2.4$                                                       & $145.1 \pm 3.0$  & $138.1 \pm 2.6$  & $139.8 \pm 2.9$  & {\color{lightgray}-} \\
             & XAI Unfaithfulness            & {\color{lightgray}0}                                                                   & $68.4 \pm 1.2$   & $48.5 \pm 0.7$   & $53.8 \pm 0.7$   &   $ 32.4 \pm 0.2     $   \\ \arrayrulecolor{lightgray} \hline
Typical      & AI Performance (inv)          & $102.9 \pm 1.3$                                                       & $113.5 \pm 1.5$  & $105.5 \pm 1.5$  & $108.5 \pm 1.6$ & {\color{lightgray}-} \\
             & XAI Unfaithfulness            & {\color{lightgray}0}                                                                     & $54.3 \pm 1.5$   & $35.1 \pm 0.5$   & $41.3 \pm 1.1$   & $26.9 \pm 0.4$             \\ \arrayrulecolor{lightgray} \hline
Outlier      & AI Performance (inv)          & $213.2 \pm 7.9 $                                                      & $231.2 \pm 8.9$  & $227.0 \pm 9.2$  & $225.1 \pm 9.0 $ & {\color{lightgray}-} \\
             & XAI Unfaithfulness            & {\color{lightgray}0}                                                                     & $107.0 \pm 3.4$  & $85.1 \pm 2.7$   & $87.7 \pm 2.1$   & $47.0 \pm 1.8$           
    \vspace{0.15cm} \\ 

\arrayrulecolor{black} \hline 
\multicolumn{2}{l}{\textbf{Heart Disease}} & \multirow{2}{*}{\begin{tabular}[c]{@{}c@{}}AI\\ Model\end{tabular}} & \multicolumn{4}{c}{XAI types}                                    \\ \cline{4-7} 
Subspace     & Metric (Acc \%, MAE \%)                &                                                                     & Global         & Subglobal      & Incremental    & Local         \\ \hline
Combined     & AI Performance     & $85.4 \pm 2.7 \%$                                                     & $69.87 \pm 6.66 \%$   & $71.46 \pm 3.06 \%$ & $70.6 \pm 4.55 \%$ & {\color{lightgray}-} \\
             & XAI Unfaithfulness            & {\color{lightgray}0}    & $18.2 \pm 0.8$  & $15.5 \pm 0.9$ & $15.7 \pm 1.1$ & $8.9 \pm 0.5 $\\ \arrayrulecolor{lightgray} \hline
Typical      & AI Performance     & $85.7 \pm 2.2 \%$    & $79.73 \pm 3.71 \%$ & $78.3 \pm 3.5 \%
$ & $77.88 \pm 2.57 \%$ & {\color{lightgray}-} \\
             & XAI Unfaithfulness            & {\color{lightgray}0}                                                                     & $16.9 \pm 1.1$ & $15.3 \pm 1.5$ & $15.3 \pm 1.6$  & $8.5 \pm 0.9$ \\ \arrayrulecolor{lightgray} \hline
Outlier      & AI Performance      & $84.9 \pm 4.7 \%$                                                     & $55.85 \pm 11.05 \%$ & $61.55 \pm 4.68 \%
$ & $60.2 \pm 7.56 \%$ & {\color{lightgray}-} \\
             & XAI Unfaithfulness            & {\color{lightgray}0}                                                                     & $20 \pm 1.7$      & $15.8 \pm 0.6$  & $16.5 \pm 0.6$ &$ 9.4 \pm 0.9$  
    \vspace{0.15cm} \\ 

\arrayrulecolor{black} \hline 
\multicolumn{2}{l}{\textbf{Auto MPG}}      & \multirow{2}{*}{\begin{tabular}[c]{@{}c@{}}AI\\ Model\end{tabular}} & \multicolumn{4}{c}{XAI types}                                    \\ \cline{4-7} 
Subspace     & Metric (MAE mpg)                &                                                                     & Global         & Subglobal      & Incremental    & Local         \\ \hline
Combined     & AI Performance (inv)             & $2.65 \pm 0.3$                                                        & $3.25 \pm 0.32$  & $2.78 \pm 0.28$  & $2.84 \pm 0.34$  & {\color{lightgray}-} \\
             & XAI Unfaithfulness            & {\color{lightgray}0}                                                                     &$1.91 \pm 0.16$  &$ 1.08 \pm 0.05$  & $1.11 \pm 0.09$  & $0.53 \pm 0.04$ \\ \arrayrulecolor{lightgray} \hline
Typical      & AI Performance (inv)             & $3.37 \pm 0.42 $                                                      & $3.92 \pm 0.49 $ & $3.59 \pm 0.51$  & $3.64 \pm 0.53$  & {\color{lightgray}-} \\
             & XAI Unfaithfulness            & {\color{lightgray}0}                                                                     & $2.17 \pm 0.25 $ & $1.32 \pm 0.14 $ & $1.36 \pm 0.14$  & $0.83 \pm 0.13$ \\ \arrayrulecolor{lightgray} \hline 
Outlier      & AI Performance (inv)             & $1.93 \pm 0.58$                                                       &$ 2.58 \pm 0.30 $  & $1.97 \pm 0.49$  & $2.03 \pm 0.46$  & {\color{lightgray}-} \\
             & XAI Unfaithfulness            & {\color{lightgray}0}                                                                     & $1.69 \pm 0.30$   & $0.87 \pm 0.17$  & $0.90 \pm 0.21$   & $0.23 \pm 0.05$ \\
\arrayrulecolor{black}  \hline
\end{tabular}
\label{tab:kfolds}
\end{table*}

\begin{figure*}[t]
    \centering
    \includegraphics[width=1.0\textwidth]{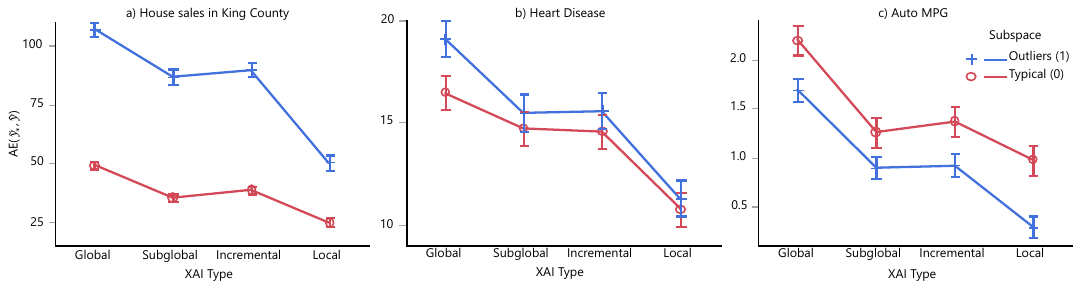}
    \caption{
    Results of modeling study showing the unfaithfulness of each explanation type calculated as absolute error (AE) between the AI Explainer estimation and AI System prediction AE($\tilde{y}$, $\hat{y}$) across three prediction tasks with different datasets: a) House Sales in King County~\cite{kingcounty}, b) Heart Disease~\cite{misc_heart_disease_45}, c) Auto MPG~\cite{misc_auto_mpg_9}.
    Global explanations are least faithful, Local explanations most faithful, and Subglobal and Incremental explanations have similar faithfulness.
    The faithfulness of typical or outlier cases depends on the explainer models trained for each dataset.}
    \label{fig:results_modeling_study}
\end{figure*}

\subsubsection{Results on faithfulness of XAI types}
Fig. \ref{fig:results_modeling_study} shows the unfaithfulness of the XAI types calculated by their absolute error (AE) between the explainer $\tilde{y}$ and predictor predictions $\hat{y}$.
Note that the explanation faithfulness ($\tilde{y}$ vs. $\hat{y}$) does not measure the same thing as predictor performance ($\hat{y}$ vs. $y$).
As expected, Global explanations had the worst faithfulness due to its low expressiveness (fewest factors), 
while Local explanations were the best because they were trained to be accurate to small local neighborhoods. However, they are not robust or memorable, so we expect users to not gain as much understanding from them compared to the other XAI types. 
Subglobal explanations had slightly better faithfulness than Incremental explanations since the latter had another objective of simplicity (fewer incremental factors). However, we expect this difference to be negligible in practical use by people, and the problem of memorability or cognitive load would override the small benefit of faithfulness, and we hypothesize that Subglobal explanations are less memorable and interpretable than Incremental explanations.
We evaluate these hypotheses later in the summative user study.

\subsubsection{Results on performance of XAI as glassbox explainers}
We further evaluate whether Subglobal and Incremental explanations can serve as accurate interpretable "glassbox" models.
In such cases, these models would be used for the AI prediction task, and be intrinsically interpretable, thus avoiding any unfaithfulness of surrogate explanation models.
We trained the models on the training dataset and report their performance. Since Heart Disease is a classification task, we accordingly apply logistic activation to the linear regression outputs and change the training objective to the binary cross-entropy loss. The interpretability of the factor coefficients is affected, since the sigmoid transform in logistic regression applies a nonlinear distortion on all weight. However, the directionality and the magnitude of the factors still provide more interpretability than blackbox models. 
We report the glassbox explainer performance as "AI Performance" in Table \ref{tab:kfolds}.
In summary, Subglobal and Incremental models performed better than Global models, but this is still a worse than the nonlinear AI model (random forest, in this case).

\subsubsection{Investigating explanations for multivariate attributes across subspaces} 

In Fig. \ref{fig:xai_types_2d} in Appendix \ref{sec:app_contour}, we show the 2D decision surfaces of the AI and explanation models for the House Sales dataset to demonstrate how the four different XAI types model the relationships between the AI prediction and multivariate attributes.

\subsubsection{Investigating varying subspace thresholds}
While Subglobal and Incremental explanations learn the feature space partitioning threshold automatically with the linear model tree, they can also be set with custom values to fit the explanation needs. We thus examine, in Appendix \ref{sec:app_ablation}, how selecting different partition thresholds affect the faithfulness of each subspace explainer model, how the factors change, and whether incremental factors are kept small.

\subsection{Formative User Study}
To investigate how people use the XAI types (Global, Subglobal, Incremental, Local) and identify usability issues, we conducted a formative study with 14 participants recruited from a local university. 
They were 23 years old on average (20 to 30), and 6 were female. 
All were undergraduate and graduate students from various disciplines (5 Sciences, 4 Business, 2 Engineering and Technology, 2 Arts, and 1 Healthcare).
The study was conducted virtually over a Zoom video call with screen recording, and lasted 60 minute. 
Participants were compensated with a digital payment of \$15 USD. 

\subsubsection{Method and Procedure}
We conducted the study with a \textit{within-subjects} experiment design, where each participant views multiple XAI types, so that they may directly compare among them.
The experiment apparatus and procedure are similar to the subsequent summative study, which we describe later.
To ensure that participants did not confuse between attribute names, values, factors, and partial contributions, we trained them to distinguish the columns in the tabular UI, understand how each partial contribution is calculated as a multiplication of factor (weight) and value, and verified their understanding with screening questions.
Each participant first used the Global explanation as a baseline, then 1-3 randomly selected XAI types as time permitted. For each explanation, the participant performed 3 trials of viewing an AI explanation to predict the price of a house instance. They were asked to estimate what they thought the AI System would predict based on the information provided by the explanation. The instances were chosen from the same dataset as the Modeling study, and used the same apparatus as the summative study (conducted later), with user interfaces shown in the Technical Approach section.

We used the think aloud protocol to elicit the participant's thought processes as they read and applied the explanations. The participant could ask clarification questions any time too. 
Since the participants were guided and supported by the experimenters, we do not report the performance of their estimations on the AI system.
With participant consent, we audio and screen recorded participant vocalizations and interactions with the UI.

\subsubsection{Findings}
We conducted a thematic analysis on participant behaviors and report key findings.

We note that some participants may have conflated the AI and XAI behavior, but we do not require our users to treat them as separate, since both are meant to be presented as a unified agent in the AI's user interface.
In this study, we focus on how participants interpreted each XAI type, rather than their trust or decision with respect to the AI prediction model.

\textit{a) Dynamic explanations perceived as more realistic than static, global explanations.}
Most participants preferred explanations to be dynamic rather than static like in Global explanations.
Only P13 felt that \textit{“the fixed factors of [Global] are more intuitive and similar to how many people think.”} Perhaps, she preferred rules of behaviors to be consistent and unchanging.
On the other hand, many participants appreciated the complexity of house price estimations and AI systems. 
P1 believed that the AI system \textit{"is a bit more dynamic in nature, or the equations will adjust accordingly to how much data is set into the thing."} He felt that Global explanations did not reflect the AI system well since \textit{"it would just come up with one static figure because the factors itself is consistent and doesn't change across the house type."}
P7 expected to see different factors for instances of different categories, remarking that \textit{"it's not very realistic for the factors to be the same for different house types,like factor for bathrooms is always [the same]}". 
This suggests she categorizes instances into types and expect rules to apply different for each category.
In contrast, 
P8 appreciated the adaptiveness of non-Global explanations and felt that \textit{"it's logical that the factors would change for different type of houses, ... since there might be other factors that influence the factor values for each attribute}."
Similarly, on seeing that \textit{"all the factors were the same"} for the Global explanation, P6 remarked \textit{"that might not be good."} He explained that for larger houses, the factor for Living Area \textit{"should be on a diminishing graph"}, i.e., smaller factor than for smaller houses due to smaller marginal utility of living area in an already large house.

Though less dynamic than Local explanations, participants found the subspace partitioning of Subglobal and Incremental explanations intuitive. 
P1 explains, \textit{"[Subglobal] is a lot more accurate [than Global] because it considers more things"}, referring to the two sets of factors given. 
P14 affirmed that \textit{"the additional factors [in Incremental] are helpful for the predictions in terms of accuracy"}.
P3 remarked that \textit{"[the Incremental factors] makes sense, because for bigger houses the land would cost more."} This also shows the relative understanding that P3 had to compare between subspaces, thus demonstrating the usefulness of explaining incrementally.

\textit{b) Incremental explanations perceived as more memorable and accurate than Subglobal explanations.}
Both Incremental and Subglobal explanations partition the subspaces similarly, but Incremental articulates the relationship between the two subspaces, and Subglobal treats them independently. Participants could appreciat the benefit of providing this context in Incremental explanations.
P6 liked that \textit{"[Incremental] would be more informed since you are telling the user how they're changing the factors, that there is an addition}".
P11 even stated that the consistency of Incremental made him feel assured because \textit{"not all the factors are changing, like there were more considerations being made by the explanation."}
P4 mentioned that \textit{"[Incremental] would be easier for me to remember because there are fewer numbers"} and 
P9 agreed that this is due to \textit{"rather than remembering two separate sets of factors"}. 
Furthermore, P12 believed \textit{"[Incremental] will give you more accurate values, which helps you make decisions quicker.”} Though, this is not necessarily true, and suggests a positive halo effect of better usability leading to perceived correctness. Nevertheless, it suggests that this can help boost user confidence, trust and usage of Incremental explanations.
Despite these benefits, some participants faced some usability issues.
P7 felt that \textit{"[Incremental] feels logical... but more time-consuming since it's slightly complex due to the additional factors you have to add for the calculation."}

\bgroup
\def\arraystretch{1.3}
\begin{table*}[bt]
    \small
    \caption{
    Hypotheses and findings of the summative user study regarding different dependent variables for various XAI types: Global (G), Subglobal (S), Incremental (I), Local (L).
    } \vspace{-0.25em}
    \begin{tabular}{lcccc}
    
    \toprule 
    Dependent Variable                                                                     & Metric            & Hypothesis                            & Finding                                                           & Evidence                                            \\ \hline
    \addlinespace[1.0mm]
    Decision duration                                                                      & Log(Time)         & G < I < S < L & I $\approx$ G < S 	$\approx$ L                                                                  & Fig. \ref{fig:results_understanding}a                                                \\
    \addlinespace[1.0mm]
    Explanation evocation                                                                  & $-$AE($\tilde{y}_h$ , $\tilde{y}$ )         & L < S < I < G & I $\approx$ S < L $\approx$ G                                                                 & Fig. \ref{fig:results_understanding}b  \\
    \addlinespace[1.0mm]
    Supported understanding   & $-$AE($\hat{y}_h$, $\hat{y}$)         & G < I < S < L & G < S $\approx$ L $\approx$ I                                                                & Fig. \ref{fig:results_understanding}c                                               \\
    \addlinespace[1.0mm]
    Sustained understanding                                                                & $-$AE($\hat{y}_h$ | $\tilde{y}$, $\hat{y}$)      & L < G $\approx$ S $\approx$ I & \begin{tabular}{c}G < L $\approx$ I $\approx$ S \\ (Special: G < L < S $\approx$ I)\end{tabular}          & Fig. \ref{fig:results_understanding}d                                               \\ 
    \addlinespace[1.0mm]
    Explanation recall   & $-$AE($w_h^{(0)}$, $w^{(0)}$)       & L < S < I < G & \begin{tabular}{c} L $\approx$ G $\leq$ S $\approx$ I \\ (L $\approx$ G $\leq$ I)\end{tabular} & Fig. \ref{fig:results_recall} \\
    \addlinespace[1.0mm]
    \begin{tabular}[c]{@{}l@{}}Perceived helpfulness\\ Perceived ease-of-task\end{tabular} & 7-pt Likert scale & L < G < S < I & L $\approx$ G $\approx$ S $\approx$ I          & Fig. \ref{fig:results_perception} \\
    \addlinespace[1.0mm]
    \bottomrule
    \end{tabular}
    \label{table:hypotheses}
    \vspace{-0.25em}
\end{table*}
\egroup

\subsection{Summative User Study}
We conducted a summative user study to evaluate the interpretability and memorability of each XAI type. 
We investigate how well participants understand, remember, and apply explanations to anticipate behavior for future instances.
While testing the impact on a downstream decision making task would be meaningful, it would impose experiment confounders, such as the participant's prior knowledge of the task~\cite{lim2009and}, their varying underlying utility objectives (e.g., how much they care about cheap housing)~\cite{lyu2020imma}, increased mental fatigue which limits the number of trials~\cite{abdul2020cogam}, and conflation between AI and XAI estimations.
Thus, we leave that for future work.

Next, we describe our experiment design and hypotheses, experiment apparatus, procedure, analysis and results.

\subsubsection{Experiment Design}
We designed our experiment as a 4$\times$2 factorial mixed-design experiment with primary independent variable (IV) as \textbf{XAI type} (four levels: Global, Subglobal, Incremental, Local) and secondary IV as \textbf{Subspace} segment (two levels: typical, special) to investigate if effects differ by instance type. 
XAI type was manipulated between-subjects due to the learning effect of participants sticking to one mental model of the AI Explainer after being trained on the first XAI type.
Subspace was manipulated within-subjects by selecting 100 instances from the full datasets, where we balanced 50 typical and 50 special. Each participant is tested on 30 randomly selected instances. 

We measured several objective dependent variables to evaluate explanation recall, application, and understanding:
\begin{itemize}
    \item \textit{Explanation recall} measures how accurately the participant can infer or remember each factor $w^{(r)}$ of the XAI type, by typing them out. This explicitly measures \textit{memorability}. For the $r$th attribute, given the participant's estimate $w_h^{(r})$, we calculate the lack of recall by the MAE. We asked about the factors for all instances (global), typical or special instances (subglobal). Although participants with Local explanations never see general explanations, we ask them to infer broadly.
    \item \textit{Sustained understanding (without XAI)} measures how well the participant can estimate the AI System's prediction. This is \textit{forward simulatability}~\cite{doshi2017towards}, a popular metric in XAI research and evaluations. Since we are modeling a regression problem (rather than the typical classification), we calculate this with a proxy metric for unfaithfulness with the absolute error (AE) of regression predictions, i.e., $|\hat{y}_h - \hat{y}|$.
    This measures how well the participant can apply knowledge gained from studying explanations for other instances without having seen their explanations. It measures deeper understanding than Supported understanding, which we also measure, described next.
    \item \textit{Supported understanding (with XAI)} measures how well the participant can estimate the AI System's prediction, given that he/she can view an approximation from the AI Explainer $\tilde{y}$, i.e., $\hat{y}_h | \tilde{y}$. This is similar to Sustained understanding, but easier, since the participant can leverage $\tilde{y}$ to estimate his answer.
    \item \textit{Explanation evocation} measures participant correctness to estimate the AI Explainer's estimation $\tilde{y}$. This is the forward simulatability of the AI Explainer, which we compute its reverse as $\tilde{y}_h - \tilde{y}$. Unlike Explanation recall, which directly elicits explanatory factors, this queries the participant about the explanation outcome, which implicitly evokes the explanation.
    \item \textit{Decision duration} measures how long participants spent to perform the forward simulatability task without XAI. Since duration follows a long-tail distribution, we analyzed its logarithm.
\end{itemize}
For participant convenience, we measured numeric factors and prediction estimates with sliders to give users bounds when answering, but use the same wide range to avoid priming.
Furthermore, we measured subjective opinions on \textit{Perceived helpfulness} and \textit{Perceived ease-of-task} to investigate how helpful the different XAI types were. Specifically, we asked whether the participant agreed or disagreed that:
the AI was accurate (1),
the explanation was helpful to estimate factors globally (2) and subglobally (3),
the forward simulatability tasks were easy with (4) or without (5) the explanation. 
These were measured on a 7-point Likert scale (-3 to +3).
Table \ref{table:hypotheses} summarize our hypotheses and the subsequent findings from our results and analysis, described later.

\subsubsection{Experiment Apparatus}\label{sec:apparatus}
\begin{figure*}[h]
    \centering
    \includegraphics[width=0.85\textwidth]{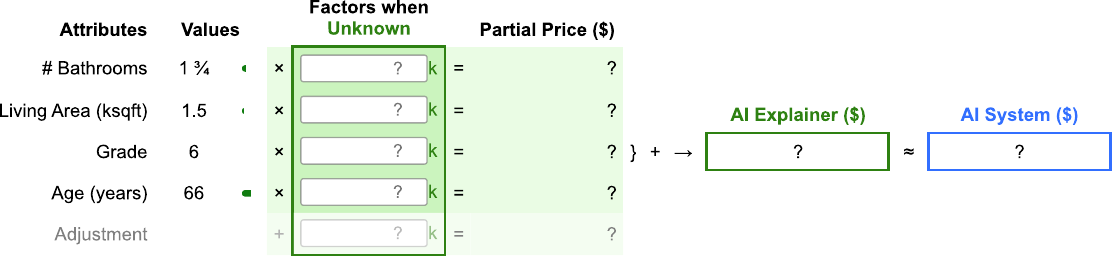}
    \caption{
    User interface (UI) during testing with factors hidden, but editable. Participants can type their own numbers to explore how the AI Explainer would compute based on various factors. This helps users to learn how factors work. Here, participants are asked to forward simulate both the AI Explainer and AI System without seeing any factor explanations.
    }
    \label{fig:ui_none}
\end{figure*}

Our user interface was inspired by the linear factors explanation interface of Poursabzi-Sangdeh et al.~\cite{poursabzi2021manipulating}, but we adapted it to distinguish the linear model explainer from the nonlinear model predictor, and extended it to support various XAI types: Global, Subglobal, Incremental, Local.
Participants saw the exact interface as shown earlier in Figs. \ref{fig:ui_global}-\ref{fig:ui_local}.
See the Appendix for the full survey that participants saw. 
We also made the UI interactive (see Fig. \ref{fig:ui_none}) to facilitate participant learning and engagement by examining how explanations and predictions depend on instance values and factors.
To improve interpretability and usability, we rounded most numbers to two significant figures, though the calculations are still done in full precision, and participants can see the precise numbers by hovering their mouse cursor. The intercept term is rounded to three significant figures, since it is a direct partial contribution component, unlike the other terms as factors.
We included green meter bars to show the relative levels of attribute values to allow participants to interpret the sense of each number.
During training, we show the \% error between the AI Explainer and AI System to make this salient and accelerate learning about explanation faithfulness.
We implemented our survey in Qualtrics and embedded the user interface.

\subsubsection{Experiment Procedure}

\begin{figure}[t]
    \centering
    \includegraphics[width=0.4\textwidth]{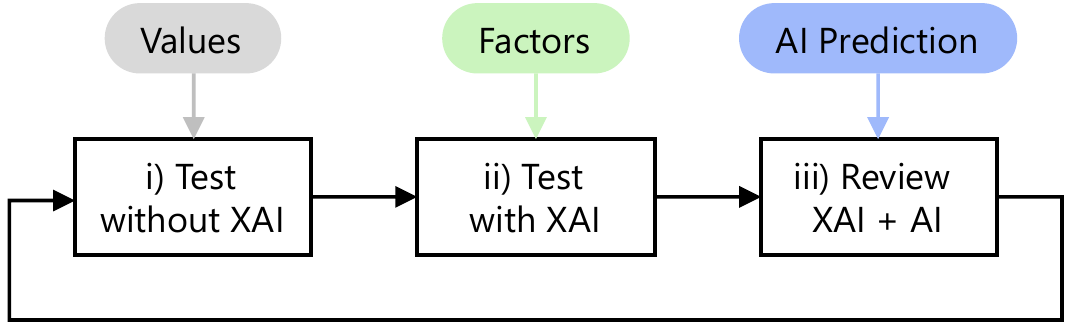}
    \caption{
    Procedure of a trial in the summative user study to evaluate explanation understanding and memorability. 
    }
    \label{fig:summative_procedure}
\end{figure}

Each participant was engaged in the following procedure:

\begin{enumerate}[label=\arabic*)]
    \item Introduction to the study (see Appendix Figs. 
    \ref{fig:intro_description}-\ref{fig:intro_prior}).
    \item Consent to participate. This study was approved by the university institutional review board (IRB).
    \item Tutorial on the AI prediction task (housing price prediction) (see Fig. \ref{fig:tutorial_basic}-\ref{fig:tutorial_explainer_system}).
    \item Tutorial on the user interface and test tasks. Different features are introduced depending on XAI type condition (see Figs. \ref{fig:tutorial_subglobal}, \ref{fig:tutorial_incremental}, and \ref{fig:tutorial_local}).
    \item Screening questions to ensure that the participant can interpret and use the explanation factors correctly (see Figs. \ref{fig:screening_global}, \ref{fig:screening_subglobal}, \ref{fig:screening_incremental}, and \ref{fig:screening_local}). 
    \item Forward simulatability session (see Fig. \ref{fig:summative_procedure}) of 5 trials with reflection and 25 regular trials, where each trial:
    \begin{enumerate}[label=\roman*)]
        \item On page 1 (see Fig. \ref{fig:trial_page1}), \textit{view} the user interface with only values shown and \textit{forward simulate} what the AI Explainer and AI System will output, as \textit{Explanation evocation} ($\tilde{y}_h$) and \textit{Sustained understanding} ($\hat{y}_h$), respectively. Factors and consequent calculations are hidden. The UI is interactive to allow the participant to type different factor values while attempting to estimate the AI outputs (see Fig. \ref{fig:ui_none}).
        Participants used two sliders to indicate their estimates for the AI Explainer and AI System outputs (see the bottom of Fig. \ref{fig:trial_page1}).
        To enhance the learning of the AI Explainer and AI System behavior and help participants learn to apply the factors, we posed several reflection questions (see the middle section of Fig. \ref{fig:trial_page1}). These were only asked for the first 5 trials to limit the survey duration.
        \item On page 2 (see Fig. \ref{fig:trial_page2}), additionally \textit{view} explanatory factors and AI Explainer calculations, based on XAI type condition, and \textit{forward simulate} what the AI System will predict. This measures \textit{Supported understanding}.
        \item On page 3 (see Fig. \ref{fig:trial_page3}), \textit{review} their answers with the actual AI System prediction. This frequent review allows the participant to continuously learn from his/her mistakes to pay better attention to learn the factors, and strengthen their understanding, across all conditions.
    \end{enumerate}
    \item Factors recall session (see Figs. \ref{fig:factors_g}-\ref{fig:factors_s1}), to recall the factors for a) all instances in general, b) typical instances, and c) outlier instances. No specific instance values are shown, but the participant can enter their own values to examine what factors could be suitable.
    \item Answer ratings questions on \textit{Perceived helpfulness} and \textit{Perceived ease-of-task}.
    \item Answer demographics questions.
    \item Acknowledge bonus calculations and exit.
\end{enumerate}

We provided an incentive bonus of £0.03 for each Sustained understanding task if the participant could estimate the AI System prediction correctly to within 10\% relative error (max £2.70), and max of £0.15 for each Explanation recall task (3 tasks) based on the mean relative error $\varepsilon_{MRE}$ on a test set of 100 instances, calculated as £$0.15 \times (1 - \varepsilon_{MRE})$.

\subsubsection{Participants}
We recruited workers from Prolific.co, where 160 passed screening and 336 failed. 
Participants who completed the study had an median age 37 years old (26 to 81), and were 35\% female. 
Participants completed the survey in a median time of 96 min,
and were compensated with a base of £9.00 and Median bonus of £1.20 (£0 to £2.88).

\begin{table*}[t]
\small
\caption{
Statistical analysis of responses due to effects (one per row), as linear mixed effects models with random effects, fixed effects, and their interaction effect. $F$ and $p$ values indicate ANOVA tests and $R^2$ indicate model goodness-of-fit.
}
\vspace{-0.1cm}
\begin{tabular}{llrrr}
\hline
Response & \begin{tabular}[c]{@{}l@{}}Linear Effects Model \\ 
(Participants as random effects)\end{tabular} & F     & p\textgreater{}F & $R^2$ \\ 
\hline

\multirow{5}{*}{\begin{tabular}[c]{@{}l@{}}Explanation evocation\end{tabular}} 
& XAI Type $+$                          & 13.3                      & \textless{}.0001         & .483 \\
& Subspace $+$         & 139.8                     & \textless{}.0001         & \\
& XAI Type $\times$ Subspace                           & 20.0                     & \textless{}.0001         & \\
& Trial ID $+$                        & 127.8                     & \textless{}.0001         & \\
& Instance ID                 & 7.5                     & \textless{}.0001         & \\
\arrayrulecolor{lightgray} \hline

\multirow{9}{*}{\begin{tabular}[c]{@{}l@{}}Supported and Sustained \\ Understanding\end{tabular}} 
& XAI Type $+$                          & 18.2                      & \textless{}.0001         & .386 \\
& Subspace $+$         & 242.0                     & \textless{}.0001         & \\
& Test with XAI $+$                              & 912.1                     & \textless{}.0001         & \\
& XAI Type $\times$ Subspace                        & 12.2                     & \textless{}.0001         & \\ & Test with XAI $\times$ Subspace $+$                 & 28.5                     & \textless{}.0001         & \\ & XAI Type $\times$ Test with XAI $+$                        & 22.6                     & \textless{}.0001         & \\

& XAI Type $\times$ Test with XAI $\times$ Subspace $+$                  & 4.5                      & .0039               & \\ & Trial ID $+$                        & 94.6                    & \textless{}.0001         & \\
& Instance ID $+$                        & 15.2                     & \textless{}.0001         & \\
\arrayrulecolor{lightgray} \hline

\multirow{5}{*}{\begin{tabular}[c]{@{}l@{}}Log(Task time w/o XAI)\end{tabular}} 
& XAI Type $+$                 & {5.8}  & {.0006} & .628 \\
& Subspace $+$ & {24.8}  & {\textless{}.0001} & \\
& XAI Type $\times$ Subspace $+$                   & {5.8}  & {.0006} & \\
& Trial ID $+$                                    & 4854.0                      & \textless{}.0001         & \\
& Instance ID                                            & 1.5                     & .0016                    & \\
\arrayrulecolor{black} \hline

\multirow{3}{*}{\begin{tabular}[c]{@{}l@{}}Explanation recall\end{tabular}} 
& XAI Type $+$                 & {8.6}  & {\textless{}.0001} & .763 \\
& Subspace $+$ & {34.5}  & {\textless{}.0001} & \\
& {\color{lightgray}XAI Type $\times$ Subspace}                 & {\color{lightgray}1.7}  & {\color{lightgray}n.s.} & \\
\arrayrulecolor{black} \hline

\multirow{1}{*}{\begin{tabular}[c]{@{}l@{}}Perceived helpfulness\end{tabular}} 
& {\color{lightgray}XAI Type}                 & {\color{lightgray}1.7}  & {\color{lightgray}n.s.} & .977 \\
\arrayrulecolor{lightgray} \hline
\multirow{1}{*}{\begin{tabular}[c]{@{}l@{}}Perceived ease-of-task\end{tabular}} 
& {\color{lightgray}XAI Type}                 & {\color{lightgray}0.1}  & {\color{lightgray}n.s.} & .863 \\
\arrayrulecolor{black} \hline

\end{tabular}
\label{table:statModelDetails}
\vspace{-0.1cm}
\end{table*}

\subsubsection{Statistical Analysis}

We performed a linear mixed effects model fit on each dependent variable as the response, XAI type and Subspace, along with other confounding variables as fixed effects, some interaction effects among the factors, and Participant as random effect. See Table \ref{table:statModelDetails} for details. 
Note that Supported understanding and Sustained understanding are calculated from the same measure, forward simulatability, but differ only by when the task was posed, before and after showing XAI, respectively. Since they share the dependent variable, we analyze these responses with a single linear mixed effects model with Test with XAI to distinguish between the two types of understanding.

The model fit was good for Log(Task time) ($R^2 = .628$) indicating that task time depended much on fixed effects XAI type, subspace, trial sequence specific test instance, and the participant random effect.
The model fit for Explanation recall was good ($R^2 = .763$), indicating that it was influenced much by two factors (XAI type and Subspace).
The model fits for Perceived helpfulness and ease-of-task were also very good ($R^2 = .977$ and .863, respectively), though there were no significant effect due to XAI type, suggesting high variance based on participant individual effect.
The model fit was slightly poorer for Explanation evocation ($R^2 = .483$), due to the difficulty to recall the explanation factors and apply weight sum arithmetic to estimate the AI Explainer's prediction $\tilde{y}_h$, leading to increased variance in participant performance.
The model fit for Understanding was somewhat low ($R^2 = .386$), because estimating the AI System's prediction regardless of explanation ($\hat{y}_h$ and $\hat{y}_h | \tilde{y}$) are even more difficult and uncertain than estimating $\tilde{y}_h$. When analyzing Supported and Sustained understanding in separate models instead of a larger model with the ``Test with XAI" factor, we obtained better model fits ($R^2 = .415$ and .479, respectively) which is similar as for Explanation evocation, but this does not properly account for viewing XAI as a causal factor.

\subsubsection{Quantitative Results}

Table \ref{table:statModelDetails} summarizes the model fits in terms of our hypotheses.
We describe our results in terms of each dependent variable and summarize the findings with respect to each XAI type. All fixed effects reported are very statistically significant (p<.0001), and describe specific comparisons based on contrast tests.
We discuss
i) how well participants could estimate the AI Explainer and AI System predictions given different XAI types based on the Forward simulatability trials, 
ii) their ability to recall the factors of each XAI type in the Factors recall session, and
iii) their perceptions on XAI helpfulness and ease-of-task.

\begin{figure*}[h]
    \centering
    \includegraphics[width=0.85\textwidth]{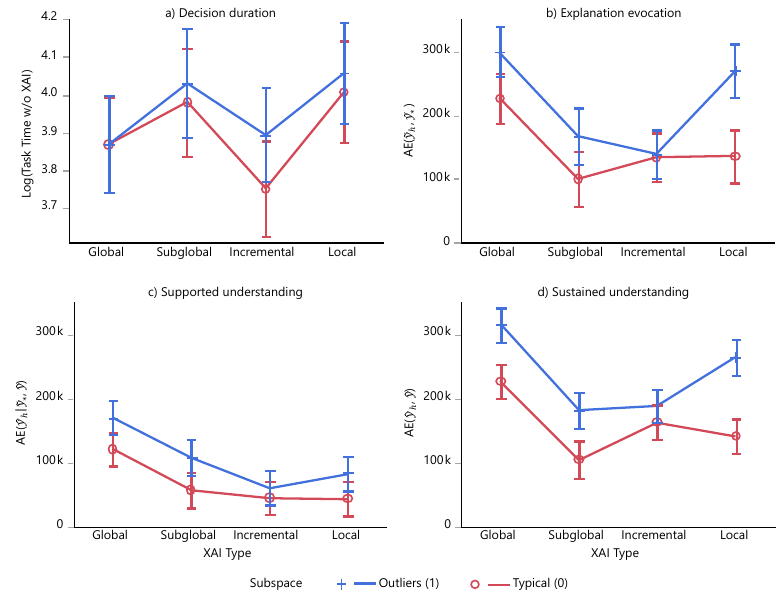}
    \caption{
    Results from forward simulatability trials to estimate the AI Explainer and AI System outputs without viewing explanations (b, d), and estimate the AI System output with explanation with timing (a, c).
    Error bars indicate 90\% confidence interval.
    }
    \label{fig:results_understanding}
\end{figure*}

\textit{Forward simulatability tasks.}
Participant performance varied across XAI types, but were generally poorer for special than typical cases (p<.0001).
See Fig. \ref{fig:results_understanding}.
Next, we discuss specific effects 
and interpret their effect sizes\footnote{Due to how variance is calculated in linear mixed effects models~\cite{rights2019quantifying}, there is no agreed way to calculate standardized effect sizes for fixed main or interaction effects. Hence, we report unstandardized effect sizes, which are the raw differences of the response variables. These are adequate to convey the practical significance in application-specific contexts and interpret their practical meaningfulness~\cite{dragicevic2018can}.}.

\begin{enumerate}[label=\alph*)]
    \item \textit{Decision duration:}
    Participants who were trained on Global or Incremental explanations were 1.19 (95\% CI: 1.10 to 1.30) 
    times\footnote{Calculated by inverse transforming $\Delta$Log(Time), i.e., $\exp(\log(t_2) - \log(t_1)) = \exp(\log(t_2/t_1)) = t_2/t_1$.} faster (M\textsubscript{G,I} = 46.8s vs. M\textsubscript{S,L} = 55.7s) at determining the AI System's output than those trained on Subglobal and Local explanations (Contrast test: $\Delta$Log(Time) $ = 0.173 \pm 0.045$, p$<$.0001).
    \item \textit{Explanation evocation:}
    Participants with Incremental or Subglobal explanations were more accurate in estimating the AI Explainer's output by \$97.5k $\pm$ \$34.6k (95\% CI) than those with Global or Local explanations (M\textsubscript{S,I} = \$134.2k vs. M\textsubscript{G,L} = \$231.6k, contrast test p$<$.0001). Given the average house price of \$589.8k, this is 16.5\% lower error.
    \item \textit{Supported understanding:}
    Participants who viewed Global explanations were worst (highest AE) at estimating the AI System output by \$79.1k $\pm$ \$13.0k (95\% CI), with 13.4\% lower error, \textit{even after} viewing the AI Explainer's output than those who viewed Subglobal, Incremental, or Local (M\textsubscript{G} = \$145.1k vs. M\textsubscript{S,I,L} = \$66.0k, contrast test p$<$.0001). 
    \item \textit{Sustained understanding:}
    The trends in participant performance was similar as for Supported understanding, but were worse due to the increased difficulty of estimating without first viewing AI explanations.
    Participants who were trained on Global explanations were worst (highest AE) by \$97.3k $\pm$ \$13.0k (95\% CI), with 16.5\% lower error, compared to those trained on other explanation types (M\textsubscript{G} = \$270.1k vs. M\textsubscript{S,I,,L} = \$172.6k, contrast test p$<$.0001). 
    Furthermore, for Special cases, participants trained on Incremental and Subglobal explanations were better by \$104.6k $\pm$ \$13.0k, with 17.7\% less error, than those trained on Local explanations (M\textsubscript{S,I} = \$104.6k vs. M\textsubscript{G,L} = \$292.8k, contrast test p$<$.0001); this suggests that subspace explanations help users to better understand special cases.
    
\end{enumerate}

\begin{figure}[t]
    \centering
    \includegraphics[width=0.48\textwidth]{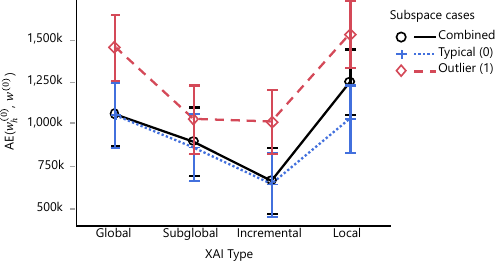}
    \caption{
    Results of explanation recall of the intercept term $w^{(0)}$ for the global and subglobal test sessions.
    }
    \label{fig:results_recall}
\end{figure}

\textit{Explanation recall task.}
We analyzed how well participants could recall or infer each factor for any instance in general (globally) or for typical or special cases (subglobally). 
While recall for most factors across global/subglobal were not significantly different, the recall for the explanation intercept term $w^{(0)}$ was notable.
Fig. \ref{fig:results_recall} shows the results of recalling $w^{(0)}$ for factor recall sessions of any, typical and outlier cases.
Participants recalled factors from Incremental explanations significantly better (lower AE) by \$456k $\pm$ \$185k (95\% CI) than from Global and Local explanations (contrast test p$<$.0001), which is practically significant compared to the intercept terms $-$\$1,040k (Combined), $-$\$697k (Typical), $-$\$1,660k (Outliers). Furthermore, though recalling Incremental factors was slightly better than of Subglobal factors, this was not significant (p = n.s.). 

\begin{figure}[t]
    \centering
    \includegraphics[width=0.35\textwidth]{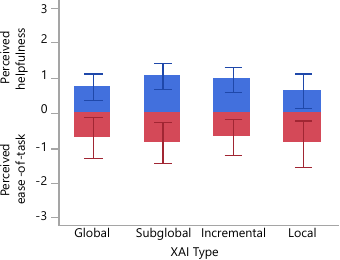}
    \vspace{-0.25cm}
    \caption{
    Results of Perceived helpfulness of AI and XAI, and Perceived ease-of-task without XAI on a 7-point Likert scale from Strongly Disagree (-3) to Strongly Agree (+3).  
    Error bars indicate 90\% confidence interval.
    }
    \label{fig:results_perception}
    \vspace{-0.15cm}
\end{figure}

\textit{Perception ratings.}
We had posed multiple questions on Perceived helpfulness and Perceived ease-of-task, but found that all perception questions except ease-of-task without explanation were correlated. Thus, we averaged them into a Perceived helpfulness metric (Cronbach's $\alpha$ = .805).
Fig. \ref{fig:results_perception}b summarizes the results of the perception measures.
Participants perceived all XAI types as somewhat helpful (M=0.84 on a -3 to +3 Likert scale), but found the forward simulatability task without XAI somewhat difficult (M=-0.75).
There were no significant differences across XAI types.

\subsubsection{Summary of results}
We now summarize our results of how each XAI type compares to others.
\begin{itemize}
    \item \textit{Global explanation} was among the fastest type due to its simplicity, but it was not the most objectively helpful for understanding due to its low faithfulness.
    \item \textit{Local explanation} supports better understanding when provided, but this understanding was not sustained for instances without explanations, since participants were unable to learn to infer factors for new cases.
    \item \textit{Subglobal explanation} supported better recall and understanding than Global or Local explanations, but participants were slow when using them to estimate what the AI System will predict.
    \item \textit{Incremental explanation} was the fastest to use (as fast as Global explanation), and best for Supported and Sustained understanding (equally good as Subglobal).
\end{itemize}

\section{Discussion}
We have introduced the paradigm of Incremental XAI, implemented its capabilities and validated its usefulness to help user understanding and recall. Here, we discuss its generalization and limitations.

\subsection{Generalizing incremental linear factors explanations}

Our implementation of Incremental explanations only had a partition along one attribute to divide instances into two subspaces. Nevertheless, since we used a tree-based partitioning method, our approach can apply to more splits and splits in multiple attributes. 
The splits can also be done on categorical attributes where each subspace can be defined by individual or a set of labels.
However, adding more splits will add complexity to the user interface and more information for users to learn and understand, especially in a short online study. Future work is needed to investigate this.
Furthermore, we had partitioned our subspaces with trees, but rules may be used to allow fewer terms instead. While tree-based subspaces cannot overlap due to the hierarchical execution of rules, rule sets for different subspaces may overlap, i.e., multiple rules with different features may be true~\cite{lakkaraju2016interpretable}. This can be mitigated with a tie-breaker~\cite{lakkaraju2016interpretable} or by using prioritized rule lists~\cite{letham2015interpretable}.

We had only investigated instances with numeric features, since we focused on training linear factors for explanation models.
To accommodate non-numeric features, such as categorical features, standard approaches to convert them, such as one-hot encoding, could be applied.
Each categorical level would be interpreted as a feature that is either present (1) or absent (0) with linear factor that is only applied when the level is present.

While our approach reduced the number of factors of the incremented weights, the base weights can also be simplified. This way, users can view an initial explanation with very few attributes, then incrementally learn new attributes for special cases or further details. This can be accomplished by applying sparsity regularization to the base factors, and a loss penalty to adding new incremental factors (which facilitates new factors if beneficial to the loss).
We had strongly limited the number of features to four shown to participants to manage their cognitive load. However, applications in machine learning could to involve about 100 features.
Incremental XAI that gradually shows more features can help users to eventually learn many features, and gain an understanding that will be highly faithful to the AI model.
This could be implemented by applying modeling on one subspace, iterating >2 steps, and regularizing against reusing features across steps. 
Further work is needed to model and evaluate on datasets with many features. Though, user testing would be challenging in lab or online studies, since learning new features is harder than adapting prior knowledge about existing features, and learning many features may not be feasible in short durations.

The partitioning of subspaces was determined in a data-driven manner with the tree model, but the factors can still be unwieldy. We had rounded the numeric factors for simplicity, but they could also be constrained as integers or multiples of integers~\cite{ustun2016supersparse}.
The split levels and factors could also be relatable~\cite{zhang2022towards} and presented verbally or narratively~\cite{riche2018data}, so that users can make sense and better remember them.

\subsection{Generalizing incremental explanations}
We have investigated incremental explanations for linear factor explanations, but argue that this can be generalized to other explanation techniques, such as generalized additive models (GAM) and rules or decision trees.
These models can be used for Global explanations by training on the full dataset, or for Subglobal or Incremental explanations by training on subspaces, or Local explanations by training on local neighboring instances.
For example, a base function could describe a quadratic trend in a feature, while an incremental function could describe a suppressing cubic effect for special cases;
or typical cases could be described with a rule of two features, while special cases could be described by substituting the second feature with a third one for a new rule.

First, we generalize to nonlinear models with independent features, such that each feature $x^{(r)}$ has a partial contribution $f^{(r)}$ to the prediction, i.e., $\tilde{y} = \sum_r{f^{(r)}}$.
We had modeled the contribution of each feature by a linear factor, i.e., $f^{(r)} = w^{(r)} x^{(r)}$
However, but this contribution could be nonlinear, i.e, $f^{(r)} = f^{(r)}(x^{(r)})$. Indeed, this matches the form of GAMs that combine nonlinear effects of features additively, i.e., $\tilde{y} = \sum_r{f^{(r)}(x^{(r)})}$.
Hence, for nonlinear models, extending Eq. \ref{eq:subglobal}, a generalized, nonlinear Subglobal explanations is
\begin{equation}
    \label{eq:subglobal2}
    \tilde{y}_s = \sum_{\varsigma}\sum_r{[x \in s_\varsigma] f_{s\varsigma}^{(r)}}
\end{equation}
where $f_{s\varsigma}^{(r)}$ is the nonlinear partial contribution of the $r$th feature in the $\varsigma$th subspace.
Similarly, extending Eq. \ref{eq:incremental}, the generalized, nonlinear Incremental explanation is
\begin{equation}
\label{eq:incremental2}
    \tilde{y}_i = \sum_{r}{\left(f_{i0}^{(r)} + \sum_{\varsigma > 0}{[x \in s_\varsigma] \Delta f_{\Delta i\varsigma}^{(r)}} \right)}
\end{equation}
where $f_{i0}$ is the base contributions of the typical explanation model, and 
${\Delta f}_{\Delta i\varsigma}$ is the incremental contributions of the $\varsigma$th special subspace explanation model.
Here we consider that the incremental difference is additive, i.e., linear. Nonlinear effects could be investigated with multiplicative interactions ($x^2 \rightarrow x^3$) or a kernel transformation.
To keep Incremental explanations simple, we can constrain the incremental contributions ${\Delta f}_{\Delta i\varsigma}$ with a sparsity regularization to reduce the number of terms, and with a smoothness regularization~\cite{abdul2020cogam} to penalize overly curvy lines.

Next, we discuss generalizing Incremental explanations to models with interaction effects, i.e., multivariate functions that involve multiple features, e.g., $f(x^{(1)},x^{(2)})$.
Common models are rules and decision trees.
While we have discussed several works to model subspaces with rules and trees, they do not support an incremental approach~\cite{lakkaraju2019faithful, natesan2020model}. 
To do so, future work could first convert any rule representation into a decision tree, compute the similarity between trees in each subspace (e.g., by calculating a graph edit distance~\cite{gao2010survey}), and minimizing the difference. However, note that rules may overlap and lead to overlapping subspaces~\cite{lakkaraju2016interpretable}.

\subsection{Scope of incremental explanations}
We evaluated Incremental explanations for the understanding and memorability of explanatory factors in AI, specifically, for users to estimate the predictions that an AI would make (forward simulatability). 
This is meant to help human \textit{cognition} toward decision making.
Further work is needed to investigate whether these lead to improvements in downstream decision making, e.g., to decide whether to accept or reject a case based on quality estimations~\cite{wang2021show}; such a study would require careful framing and incentivization to ensure that users are correctly aligned and properly motivated to the task, and avoiding the confounder of prior knowledge which can diminish the benefits of XAI~\cite{lim2009and}.
We do not propose it for \textit{perception} tasks (e.g., vision and audio) or language reasoning (NLP), since they involve innate mental processes due to stimuli or low-level skills rather than deliberate reasoning. 

Our paradigm of explanation incrementation assumes that users are novices who start with limited knowledge of the domain or AI application, thus they need to be taught gently. We do not expect Incremental explanations to be strongly beneficial for domain experts who can handle complex data and have established conventions~\cite{lim2023diagrammatization}.

Similar to Poursabzi-Sangdeh et al.~\cite{poursabzi2021manipulating}, we had evaluated Incremental explanations only for one application task of predicting housing prices. Perhaps, for applications that are less common (e.g., health diagnosis), or with critical but complicated numbers (e.g., decimals or fractions), Incremental and Subglobal explanations may still be overwhelming. Future work should validate our results across other application tasks.

\subsection{Implications of incremental explanations}
Our approach for Incremental explanations enables better learning of sparse linear factor models.
This adds to the body of work of subspace-based explanations that take a divide-and-conquer approach to partitioning instances, and explaining each subspace as similarly as possible. 
COGAM moderated the number of visual chunks in line graphs~\cite{abdul2020cogam}, and it can be made to incrementally allow more curviness to allow users to learn more details. 
GlocalX~\cite{setzu2021glocalx} provides rule explanations in detail and in aggregate by merging them.
Future work could investigate which explanation format (factors, line segments, or rules) are easier and more beneficial to learn incrementally.

Although our participants could well learn and recall Incremental explanations in our study, we acknowledge that the learning time was brief. Most learning that people do occurs over longer time periods with more repetitions. Hence, future work could deploy Incremental explanations to investigate its longitudinal benefits. We note that under longer durations, the learning of Subglobal explanations may also be improved, but perhaps less so than Incremental explanations, due to the slightly higher cognitive load.

Our approach of Incremental explanations limited the explanations to the same type (sparse linear models). However, users have diverse preferences for and usage strategies of explanations~\cite{lim2011design, lim2013evaluating}, so incremented explanations should also be diverse. For example, first provide factors, then rules. This provides users with diverse retrieval cues, which can reinforce their memory of the explanations. Future work can explore how to increment across explanation structures.

\section{Conclusion}
We have introduced Incremental XAI to help users better recall and apply explanations of AI.
This provides a set of base general factors for typical instances and sparse incremented factors for special cases.
In modeling and user studies, we found that Incremental explanations help facilitate fast understanding like Global explanations that explain generally, and are easy to recall and understand like Subglobal explanations that explain subspaces more faithfully than Global explanations. Incremental explanations are also more memorable than Local explanations, facilitating better recall and understanding performance.
This work demonstrates the importance of supporting more memorable explanations to deepen user understanding of AI for more productive interactions.

\begin{acks}
This work was supported by 
the Singapore Ministry of Education (MOE) Academic Research Fund Tier 2 T2EP20121-0040, and
the DesCartes programme supported by the National Research Foundation, Prime Minister’s Office, Singapore under its Campus for Research Excellence and Technological Enterprise (CREATE) programme,
and was carried out at the
the NUS Institute for Health Innovation and Technology (iHealthtech).
We thank Jenny Yu and Lana Xu for help on the initial user interface development.
\end{acks}

\bibliographystyle{ACM-Reference-Format}
\bibliography{main}

\clearpage
\onecolumn
\appendix
\renewcommand\thefigure{\thesection.\arabic{figure}} 
\section{Appendix}

\subsection{Decision surfaces of each explanation}
\label{sec:app_contour}

In \ref{fig:xai_types_2d}, we examine the decision surfaces of the predictor and explainer models to study how the different XAI types support linear or piecewise-linear relationships between multiple attributes and the prediction value. Based on the House Sales dataset, this provides a conceptual interpretation to the reader of 
i) how nonlinear the decision surface of the AI Predictor model is (a), and how the Local explanations cumulatively capture the nonlinearity (e); 
ii) how Subspace explanations (c) better fit the nonlinear decision surface of the AI predictor model (e) compared to the Global explanation (b); and
iii) how Incremental explanations simplify memorability by first conveying a linear relationship (d, left), then showing a partial linear segment (d, right)

\setcounter{figure}{0}  
\begin{figure*}[h]
    \centering
    \includegraphics[width=0.8\textwidth]{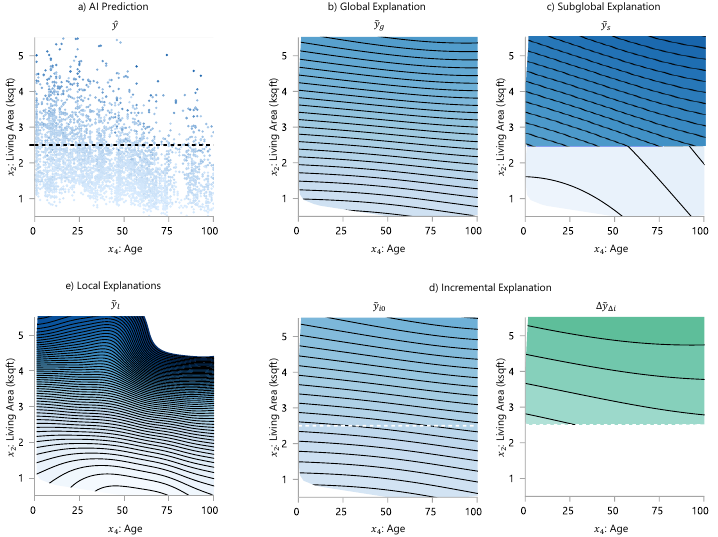}
    \caption{
    Decision surface of AI prediction model and various explanation models with two attributes (x-y axes), showing prediction output (color: darker is higher value) for the House Sales in King County, USA~\cite{kingcounty} dataset.
    Models are based on the dataset used in the modeling and user studies; showing only values for attributes 2 and 4.
    a) Scatter plot of AI System predictions for instances with various ($x_2$, $x_4$) values. Dashed line indicates threshold to split for Subglobal and Incremental explanation models.
    b) Contour plot of Global explanation showing linear slope mostly increasing in the direction of $x_2$ i.e., $w_2 > w_4$ (see factors in Fig. \ref{fig:ui_global}).
    c) Contour plot of Subglobal explanation model showing two linear models -- gentler-sloped for typical instances ($x_2 < 2.5$) and steeper-sloped for outlier instances ($x_2 \geq 2.5$).
    d) Contour plot of Incremental explanation model showing general model for all instances (Left), and incrementally-sloped model for outlier instances ($x_2 \geq 2.5$, Right). 
    e) Contour plot of accumulation of multiple Local explanation models showing non-linear surface that manifests when learning from heterogeneous local explanations.
    }
    \label{fig:xai_types_2d}
\end{figure*}

\clearpage

\subsection{Subglobal and Incremental explanations performance across varying subspace thresholds}
\label{sec:app_ablation}

Subglobal and Incremental explanations model data with two aspects: first partitioning the feature space into subspaces, then fitting linear factors for each subspace. 
Although the partition thresholds are learned automatically with the linear model tree, they may be adjusted to simplify presentation to users.
We examined how selecting different partition thresholds affect the faithfulness of each subspace explainer model, how the factors change, and whether incremental factors are kept small.

\begin{figure*}[h]
    \centering
    \includegraphics[width=1.0\textwidth]{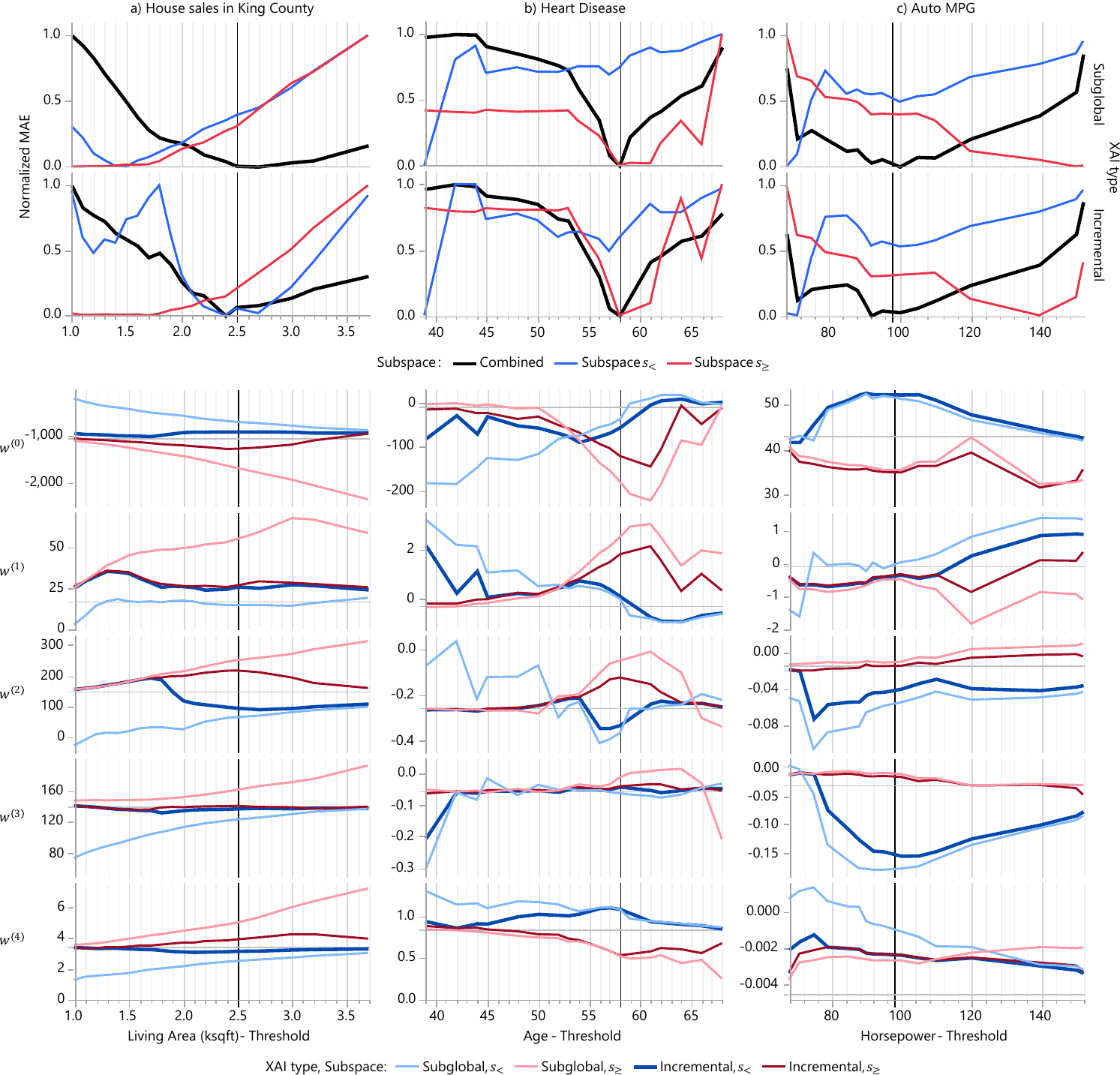}
    \caption{
    Results of XAI faithfulness and factors for Subglobal and Incremental explanations when varying subspace partitioning thresholds, for different datasets and subspaces.
    In the top row graphs, lower MAE is better, and normalized MAE is shown to clarify the curve minima. 
    In the bottom graphs, incremental factors are regularized to 0 if the weight values are the same for subspace $s_{<}$ as for $s_{\geq}$.
    Vertical black lines indicate the optimal partition threshold chosen in the modeling and user studies.
    Gray horizontal lines indicate the factors for the Global explanation.}
    \label{fig:results_modeling_thresholds}
\end{figure*}
For this experiment, we manually set the partition threshold of the feature selected by the linear model tree (Living Area $x^{(2)}$ for House Sales, Age $x^{(1)}$ for Heart Disease, Horsepower $x^{(3)}$ for Auto MPG), then trained Subglobal and Incremental models for each threshold level. We varied thresholds from the 10th to 90th percentile of feature values, see Fig. \ref{fig:results_modeling_study} for results.
We denote the subspace below the threshold as $s_{<}$ and the other as $s_{\geq}$. Note that as the threshold changes, the ratio of subspace sizes changes, so we do not label them as typical or outlier.
We calculated the Combined MAE as the sample-weighted sum of subspace MAE based on the number of instances in each subspace, then normalize each MAE metric for clarity (see Fig. \ref{fig:results_modeling_study}, top). The lowest Combined MAE coincides with the thresholds selected by the linear model tree.
Fig. \ref{fig:results_modeling_study} (bottom) shows how each linear factor $w^{(r)}$ changes with partition threshold.

As expected, the weights across XAI type for each subspace is somewhat similar (i.e., $w_{s\varsigma}^{(r)} \approx w_{i\varsigma}^{(r)}$), with differences due to different loss functions.
However, the weights of Subglobal explanations across subspaces are generally different (i.e., $w_{s<}^{(r)} \neq w_{s\geq}^{(r)}$), since each subspace is best fit with different linear models.
Conversely, the weights of Incremental explanations are sometimes similar across subspaces (i.e., ${\Delta w_{i}}^{(r)} = w_{i<}^{(r)} - w_{i\geq}^{(r)} = 0$), indicating that some incremental factors have been regularized to 0. This is seen when the dark blue and red lines overlap in Fig. \ref{fig:results_modeling_study} (bottom).
Note how the MAE may not be the lowest for such cases, so there is a trade-off to prioritize performance over simplicity (fewer Incremental factors), which is typical in XAI methods~\cite{arrieta2020explainable}
For House Sales, the incremental factors ${\Delta w_{i}}^{(1)}$ and ${\Delta w_{i}}^{(3)}$ are always almost 0, and ${\Delta w_{i}}^{(2)} = 0$ when Living Area < 1.8 ksqft.
In general, Base+Incremental factors are not identical to Global or Subglobal factors due to different training objectives.

\clearpage

\subsection{Qualtrics survey for the summative user study}
\label{sec:app_qualtrics}

\begin{figure}[h]
    \centering
    \includegraphics[width=0.9\textwidth]{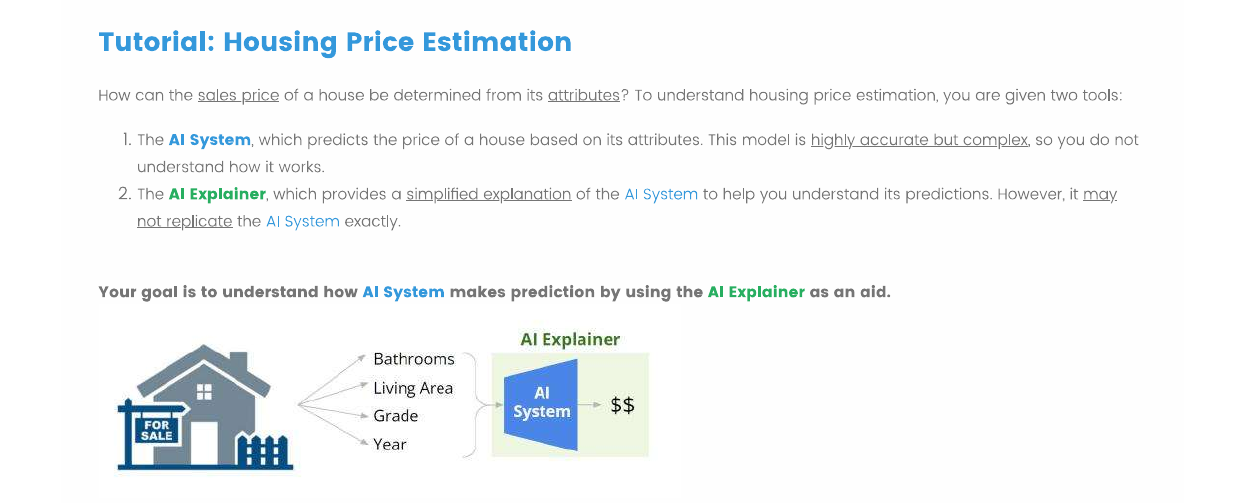}
    \caption{
   Introduction of housing price estimation.
    }
    \label{fig:intro_description}
\end{figure}

\begin{figure}[h]
    \centering
    \includegraphics[width=0.9\textwidth]{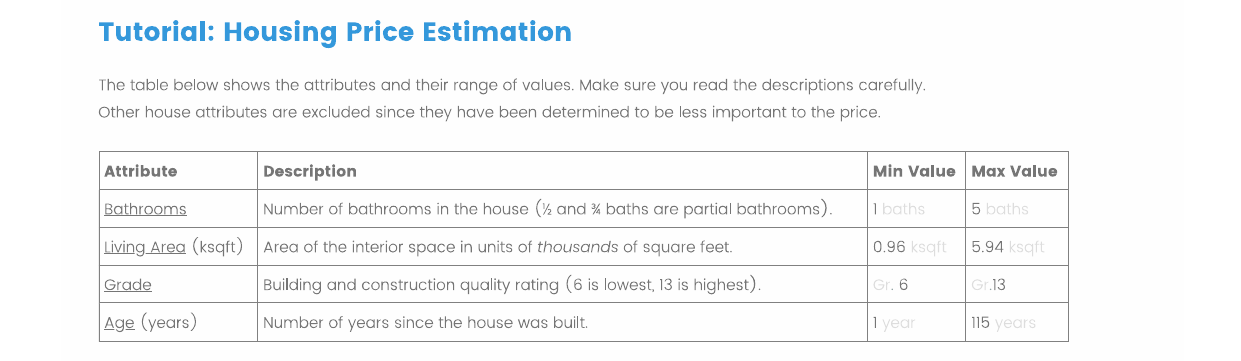}
    \caption{
   Introduction of attributes in housing price estimation.
    }
    \label{fig:intro_table}
\end{figure}

\begin{figure}[h]
    \centering
    \includegraphics[width=0.9\textwidth]{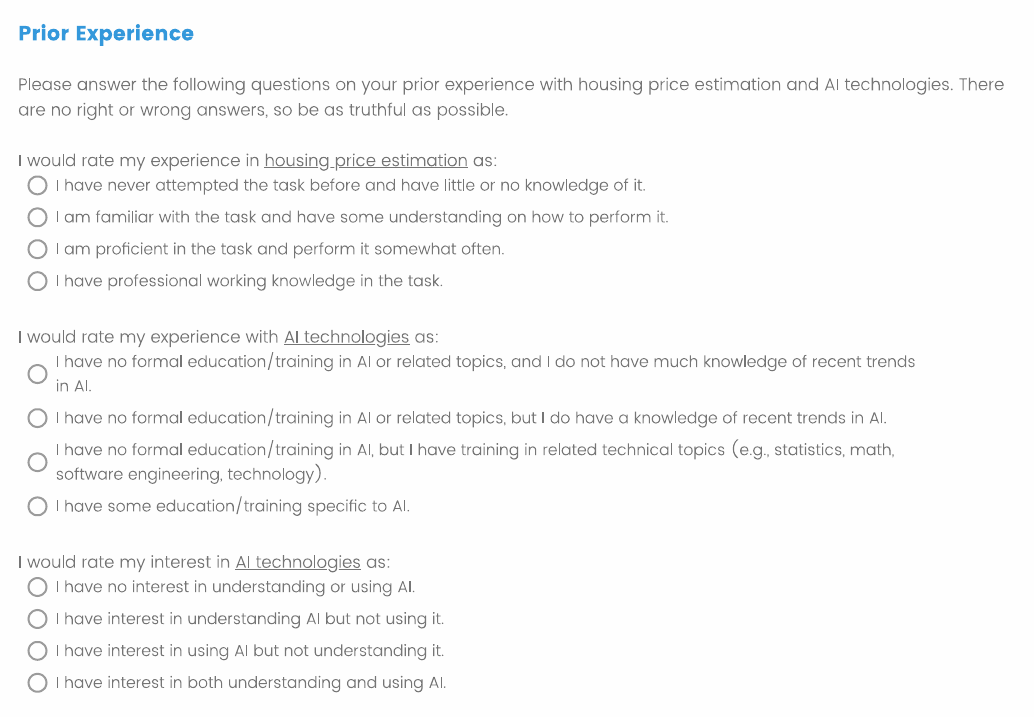}
    \caption{
   Questions on users' prior experience with housing price estimation and AI background.
    }
    \label{fig:intro_prior}
\end{figure}

\begin{figure}[h]
    \centering
    \includegraphics[width=0.9\textwidth]{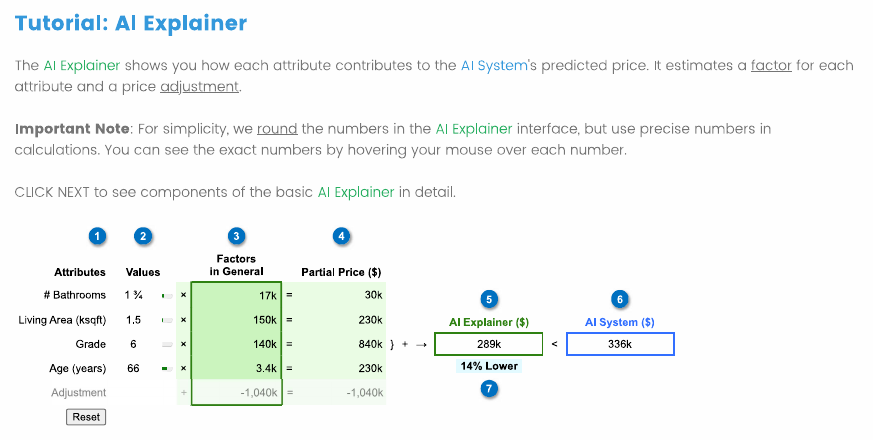}
    \caption{
   Tutorial on basic (Global) AI explanation.
    }
    \label{fig:tutorial_basic}
\end{figure}

\begin{figure}[h]
    \centering
    \includegraphics[width=0.70\textwidth]{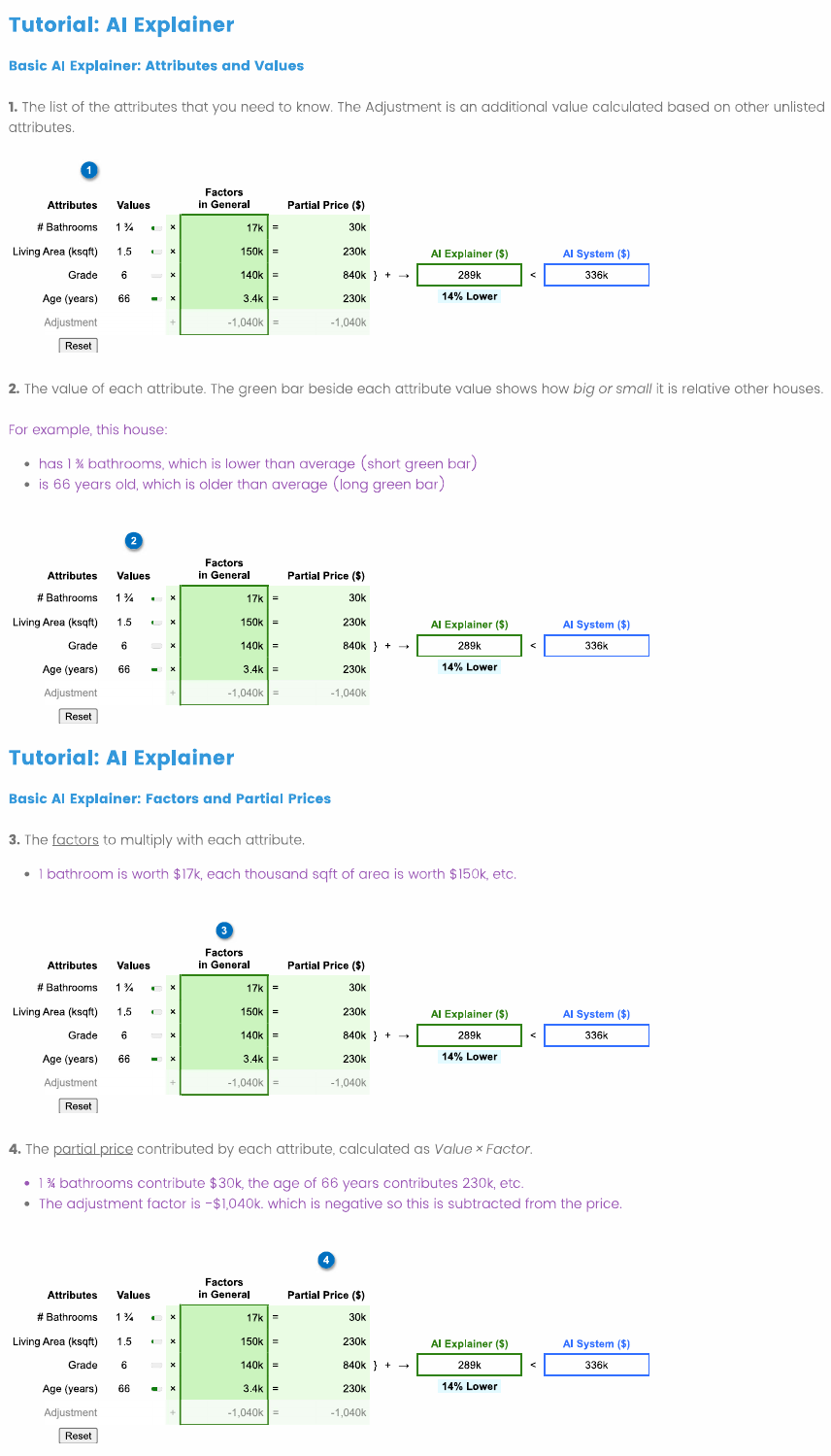}
    \caption{
   Tutorial on the attributes, values, factors, and partial prices of the explanation interface. All participants are first trained on the Global explanation since it has the simplest format. 
    }
    \label{fig:tutorial_basic_values_factors}
\end{figure}

\begin{figure}[h]
    \centering
    \includegraphics[width=0.7\textwidth]{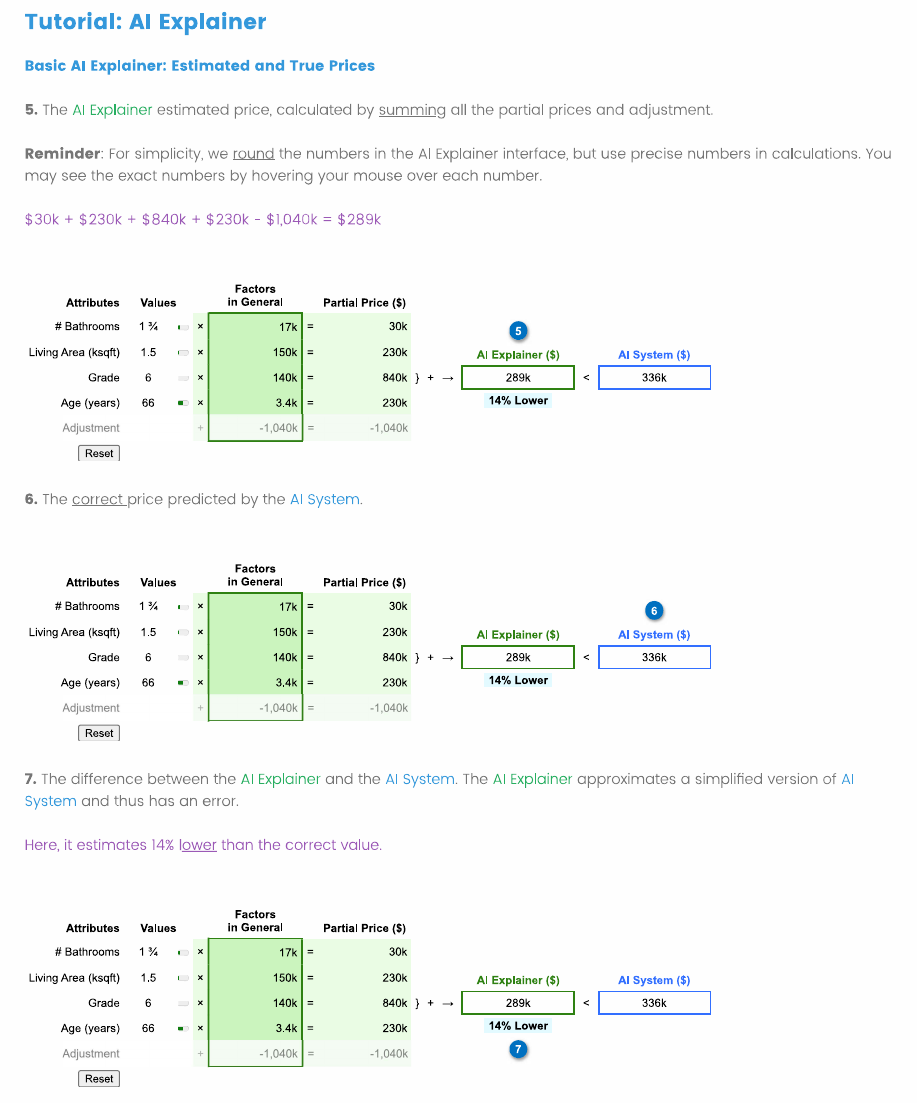}
    \caption{
   Tutorial on estimated explanation price, correct AI System price, and the percent difference.
    }
    \label{fig:tutorial_explainer_system}
\end{figure}

\begin{figure}[h]
    \centering
    \includegraphics[width=0.7\textwidth]{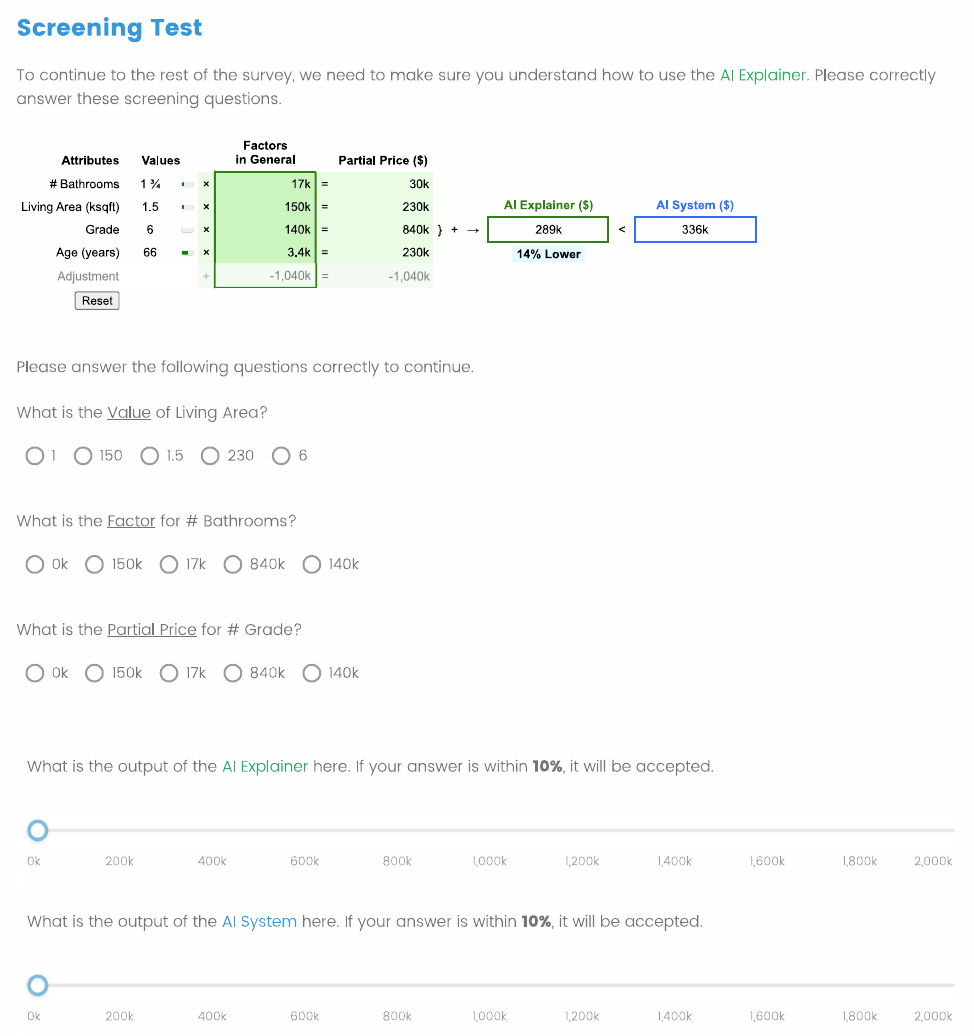}
    \caption{
    Screening questions for the Global explanation to check users’ comprehension.
    }
    \label{fig:screening_global}
\end{figure}

\begin{figure}[h]
    \centering
    \includegraphics[width=0.7\textwidth]{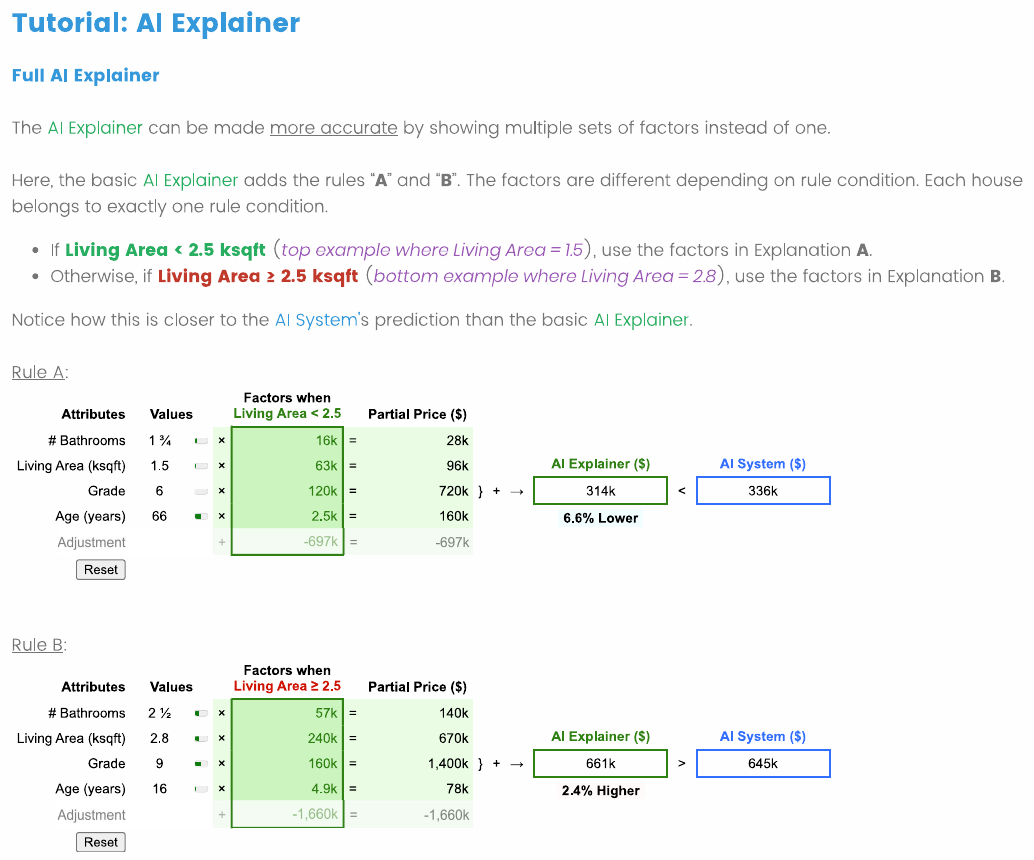}
    \caption{
    Tutorial on the Subglobal explanation.
    }
    \label{fig:tutorial_subglobal}
\end{figure}

\begin{figure}[h]
    \centering
    \includegraphics[width=0.7\textwidth]{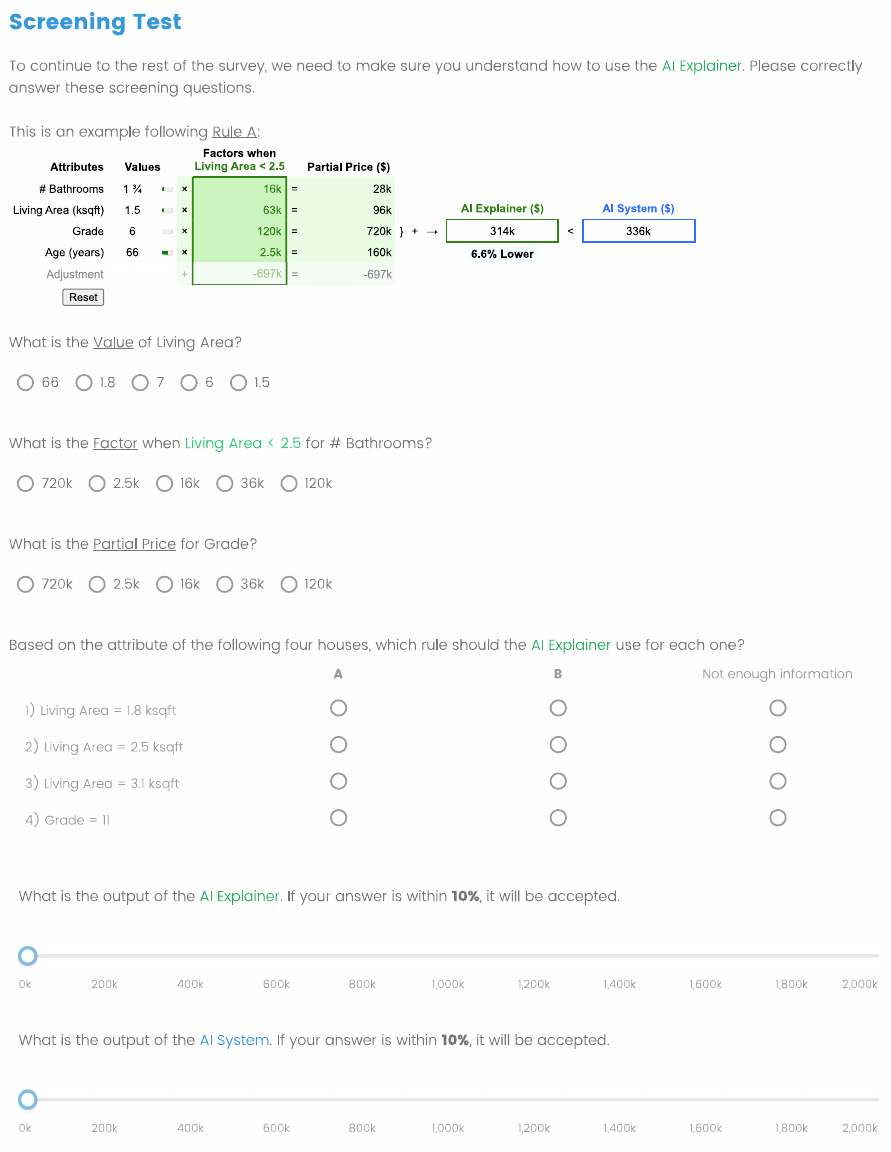}
    \caption{
    Screening questions for the Subglobal explanation to check users’ comprehension.
    }
    \label{fig:screening_subglobal}
\end{figure}

\begin{figure}[h]
    \centering
    \includegraphics[width=0.7\textwidth]{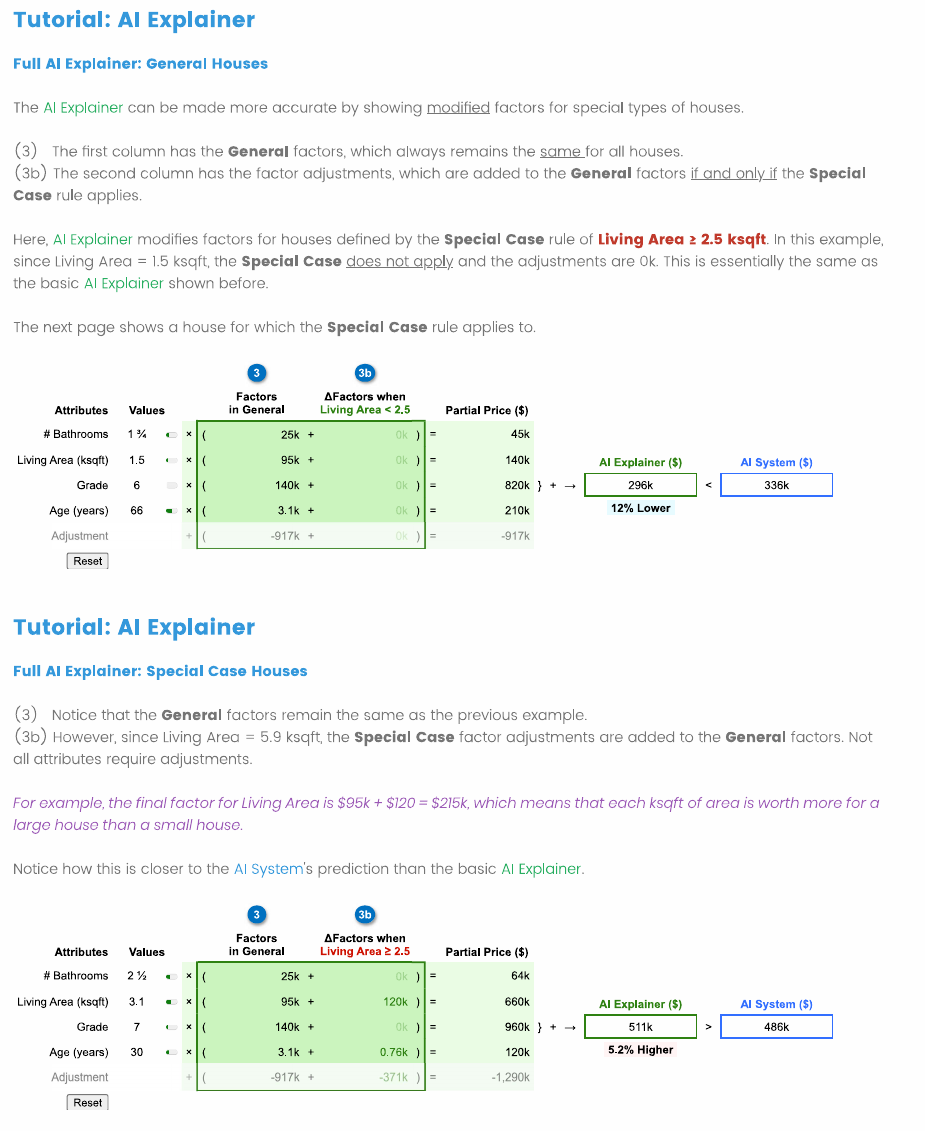}
    \caption{
    Tutorial on the Incremental explanation.
    }
    \label{fig:tutorial_incremental}
\end{figure}

\begin{figure}[h]
    \centering
    \includegraphics[width=0.7\textwidth]{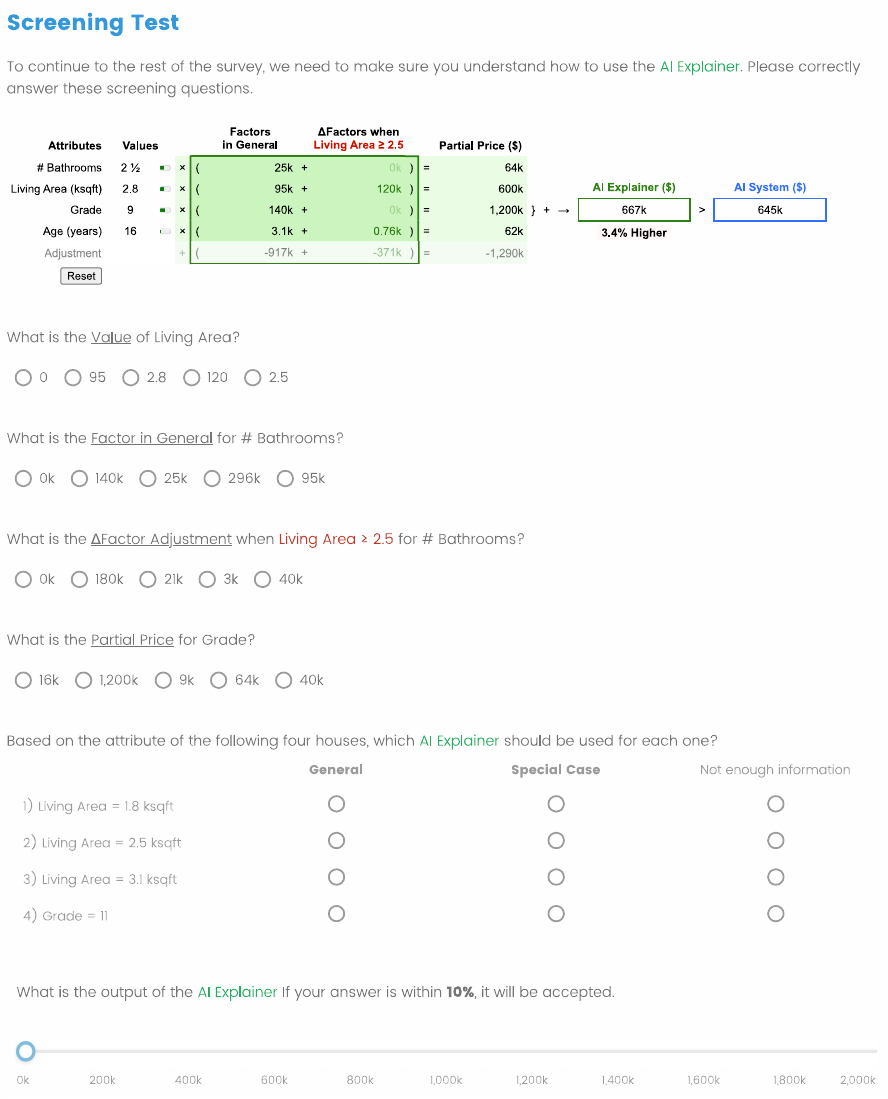}
    \caption{
    Screening questions for the Incremental explanation to check users’ comprehension.
    }
    \label{fig:screening_incremental}
\end{figure}

\begin{figure}[h]
    \centering
    \includegraphics[width=0.7\textwidth]{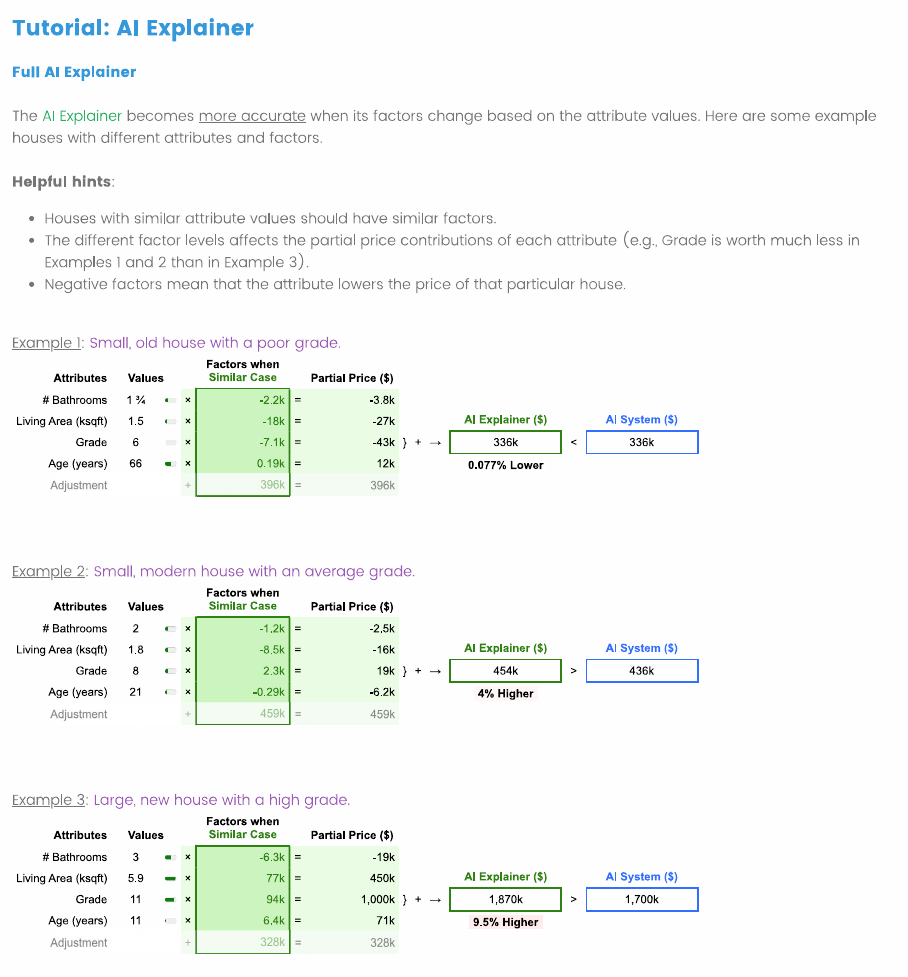}
    \caption{
    Tutorial on the Local explanation.
    }
    \label{fig:tutorial_local}
\end{figure}

\begin{figure}[h]
    \centering
    \includegraphics[width=0.7\textwidth]{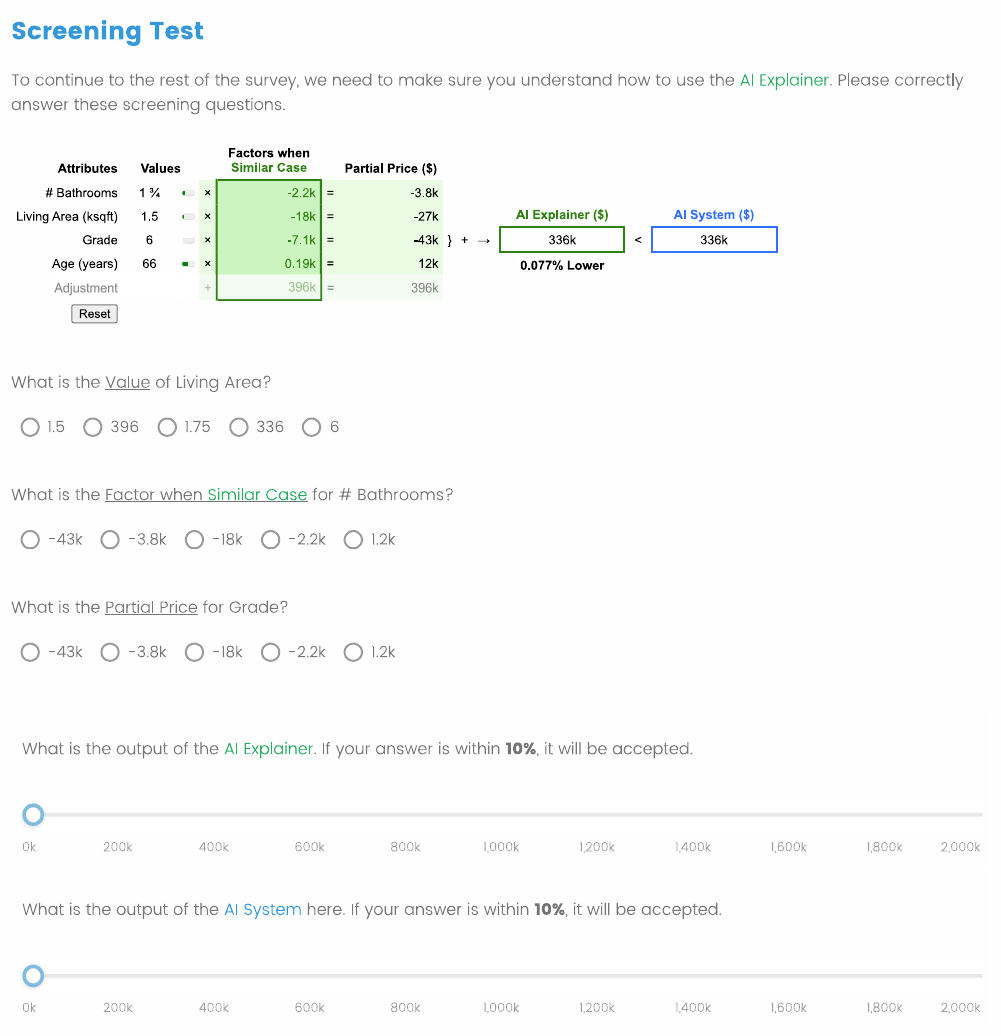}
    \caption{
    Screening questions for the Local explanation to check users’ comprehension.
    }
    \label{fig:screening_local}
\end{figure}

\begin{figure}[h]
    \centering
    \includegraphics[width=0.7\textwidth]{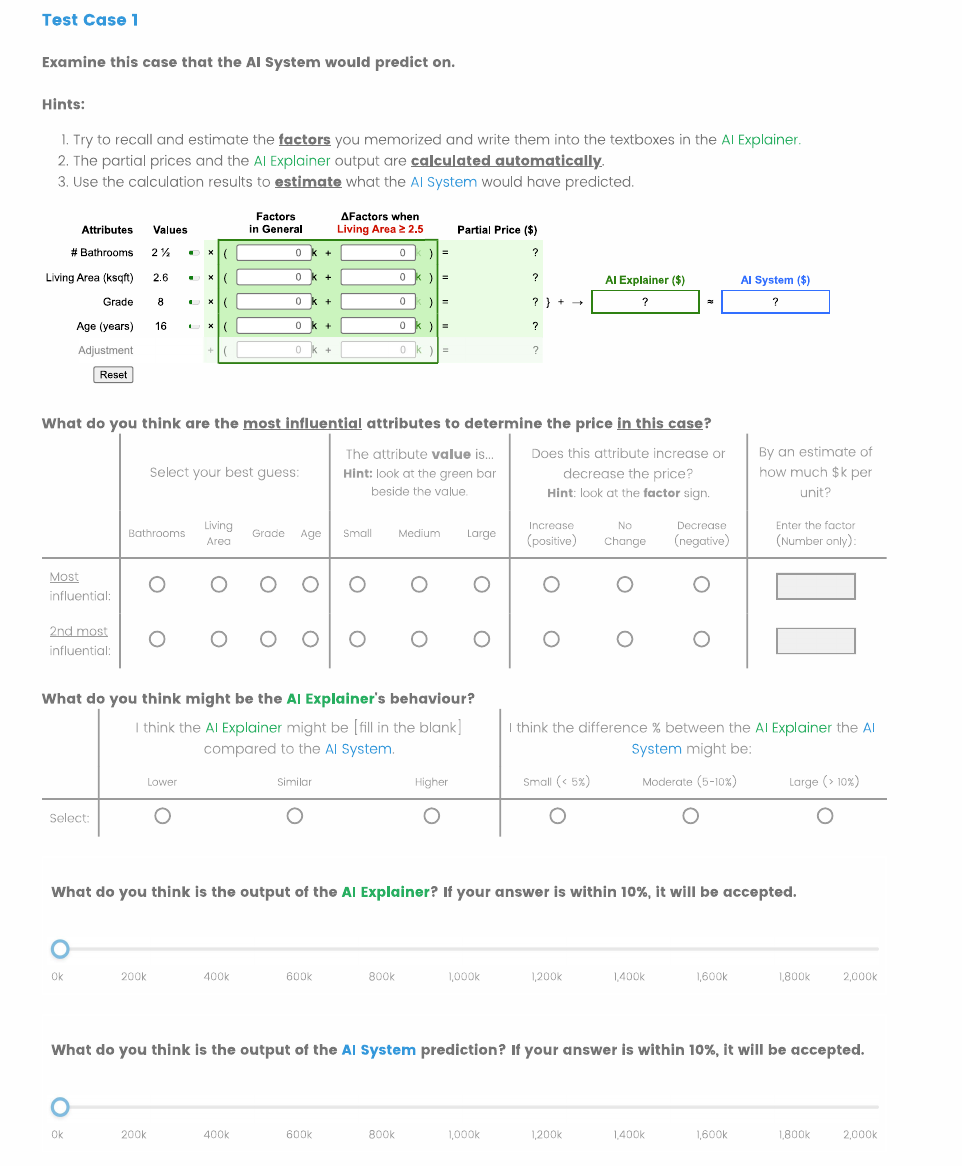}
    \caption{
    Sample of the unassisted forward simulation trial, where participants are asked to estimate the explanation and AI system outputs. The displayed UI is the Incremental condition.
    }
    \label{fig:trial_page1}
\end{figure}

\begin{figure}[h]
    \centering
    \includegraphics[width=0.7\textwidth]{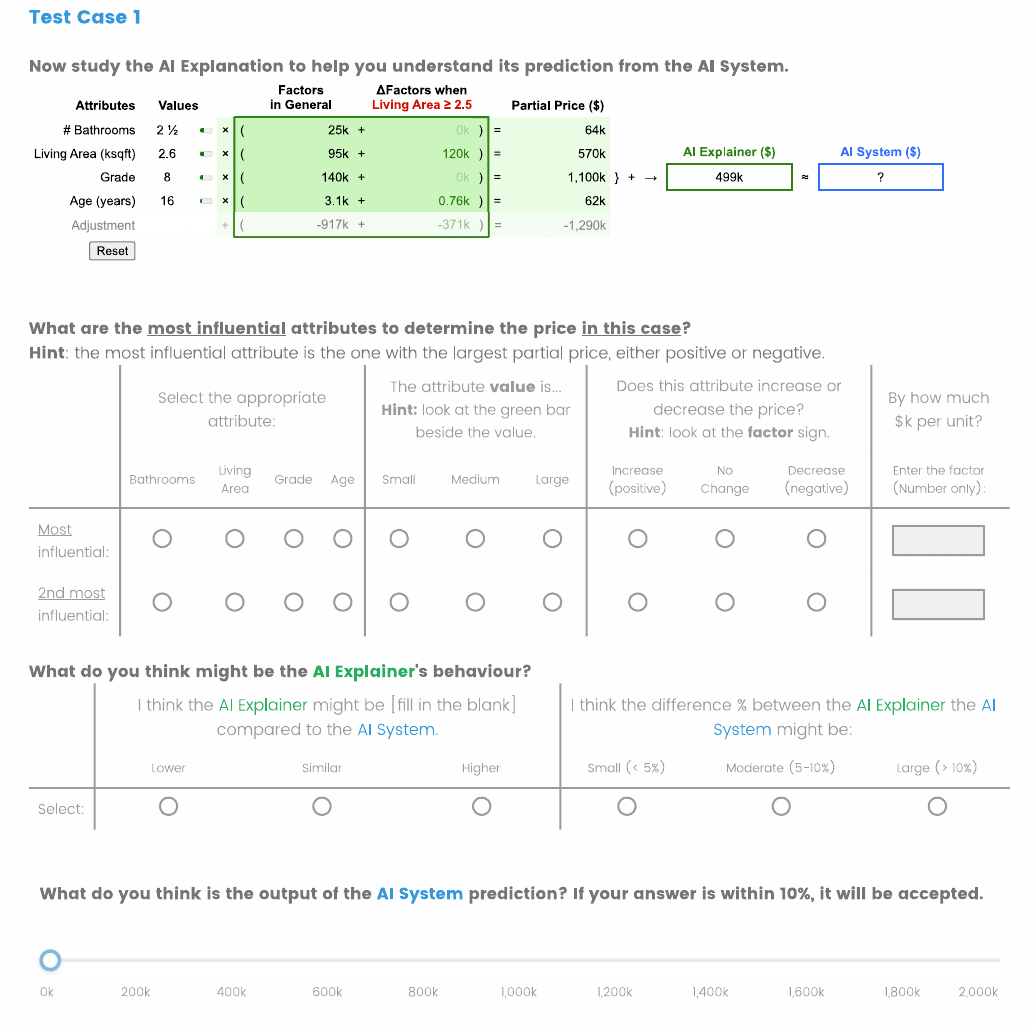}
    \caption{
    Sample of the assisted forward simulation trial, where participants are asked to estimate the AI system output based on the given explanation. The displayed UI is the Incremental condition.
    }
    \label{fig:trial_page2}
\end{figure}

\begin{figure}[h]
    \centering
    \includegraphics[width=0.7\textwidth]{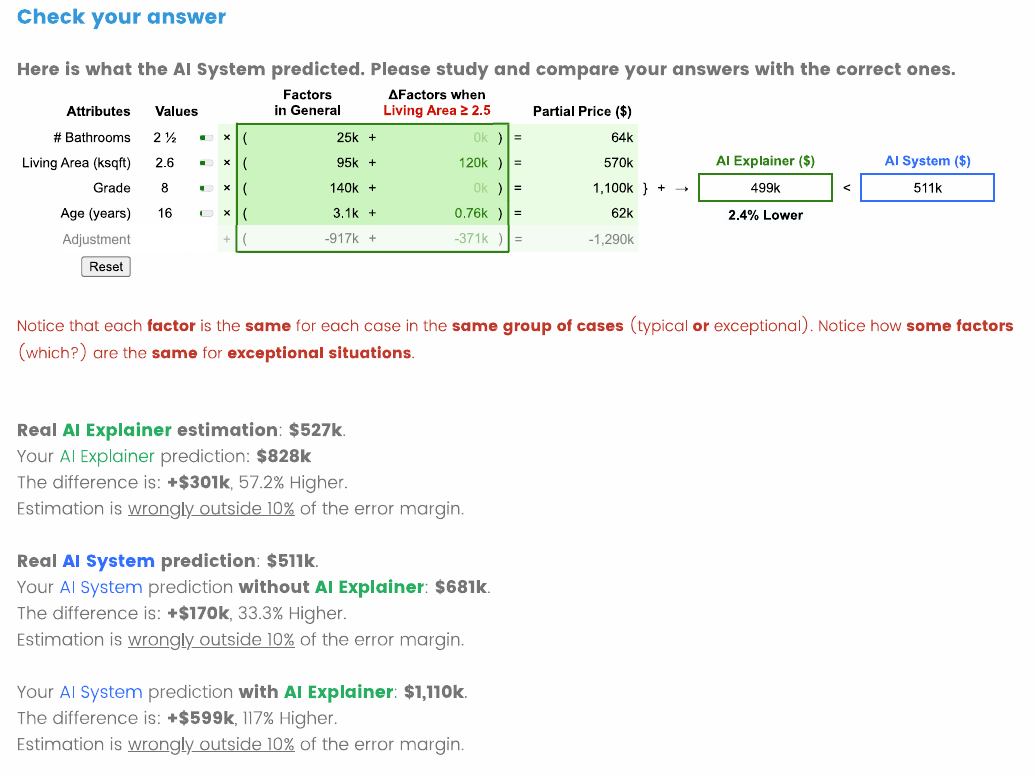}
    \caption{
    Review of performance on forward simulation trials, for participants to strengthen their understanding. The displayed UI is the Incremental condition.
    }
    \label{fig:trial_page3}
\end{figure}

\begin{figure}[h]
    \centering
    \includegraphics[width=0.7\textwidth]{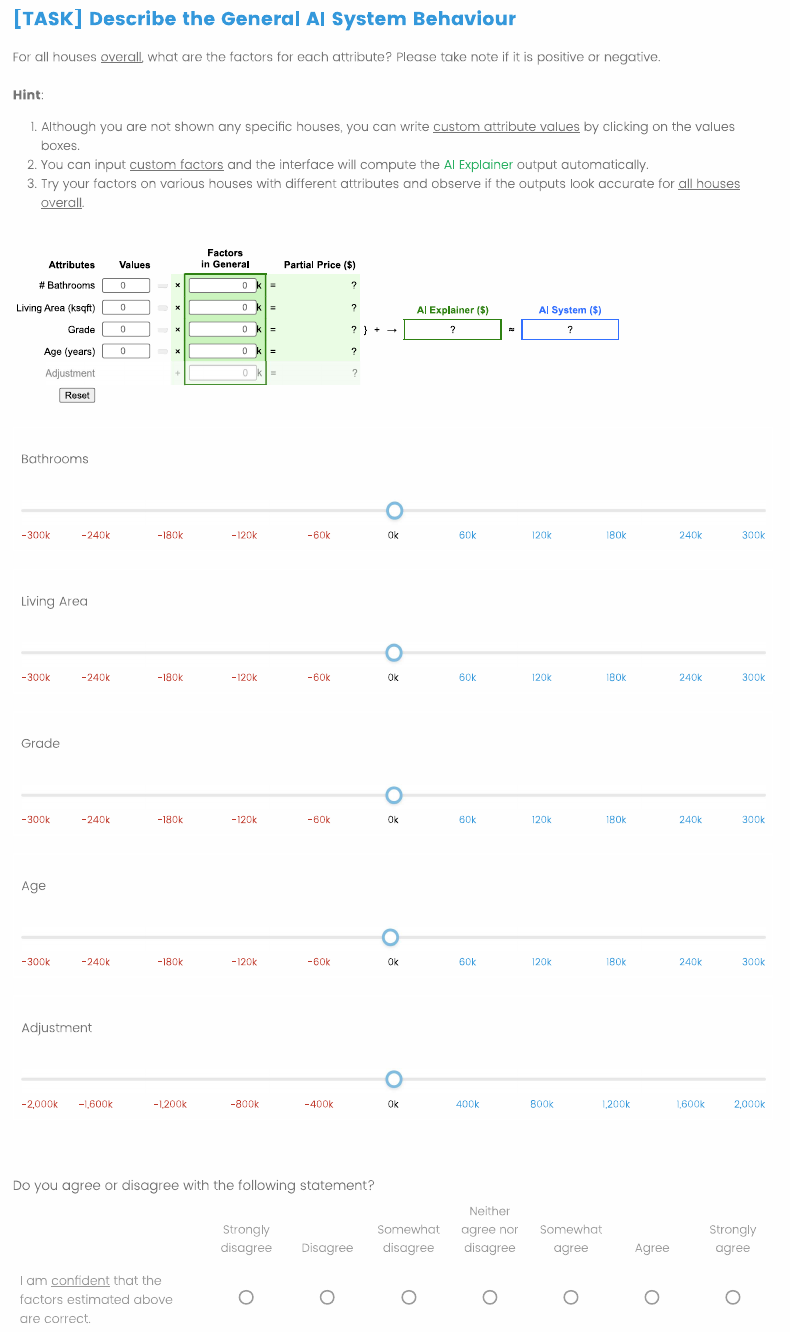}
    \caption{
    Explanation recall task for all instances overall (Global).
    }
    \label{fig:factors_g}
\end{figure}

\begin{figure}[h]
    \centering
    \includegraphics[width=0.7\textwidth]{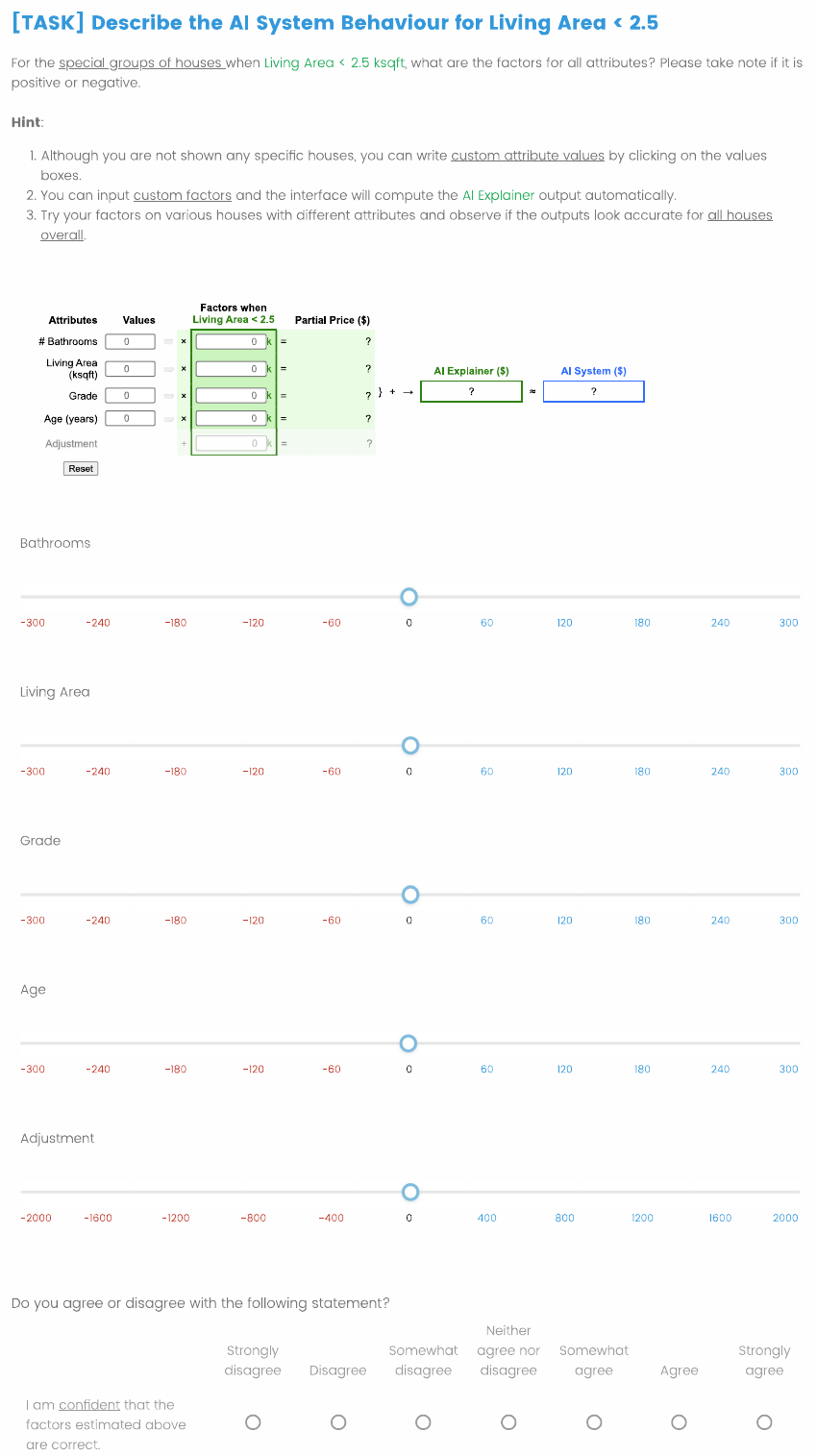}
    \caption{
    Explanation recall task for subspace Living Area \textless{} 2.5.
    }
    \label{fig:factors_s0}
\end{figure}

\begin{figure}[h]
    \centering
    \includegraphics[width=0.7\textwidth]{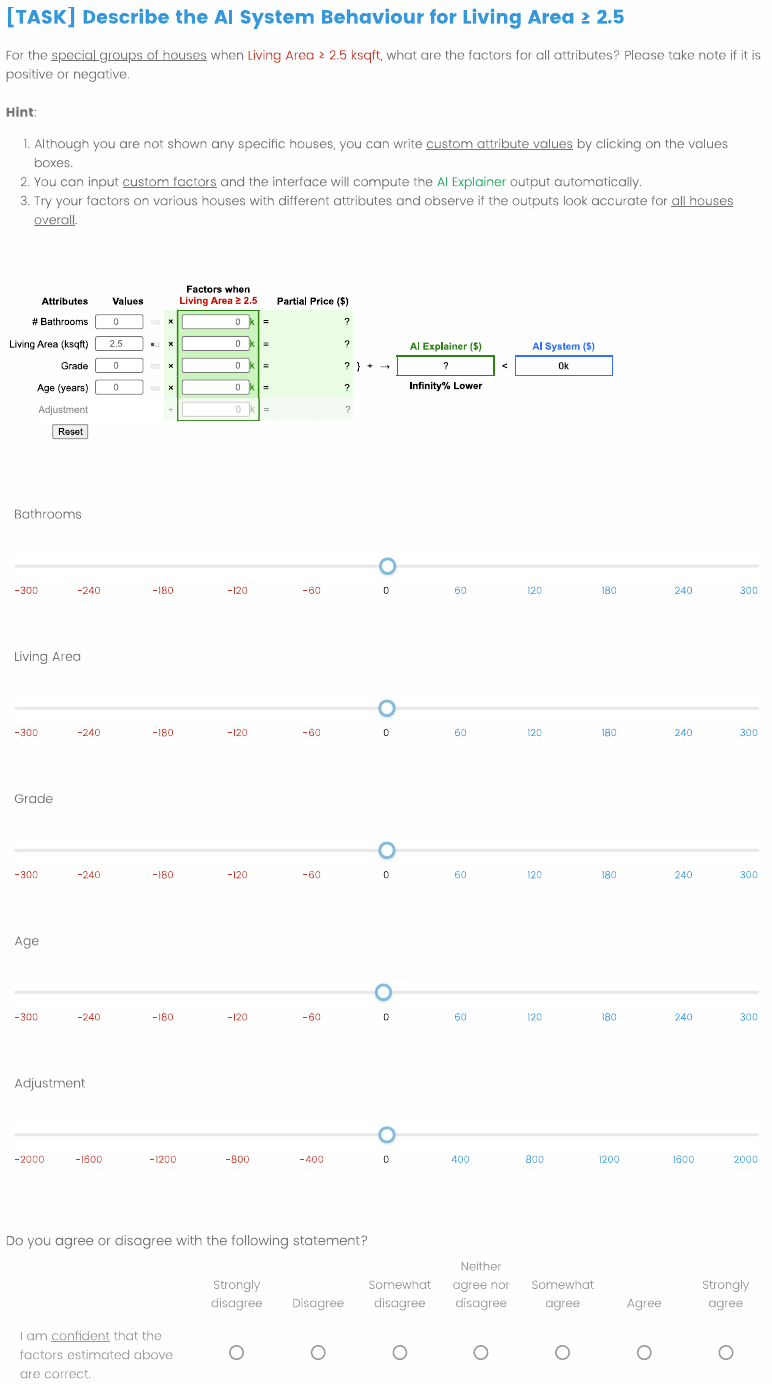}
    \caption{
    Explanation recall task for subspace Living Area $\geq$ 2.5.
    }
    \label{fig:factors_s1}
\end{figure}

Figures \ref{fig:intro_description}-\ref{fig:factors_s1} depict the Qualtrics survey workflow with each of the four XAI conditions for the summative user study.

\end{document}